\documentclass[aps,twocolumn,showpacs, nofootinbib]{revtex4}

\usepackage{wallpaper}
\usepackage{dcolumn}
\usepackage{bm}
\usepackage{pdfpages}
\usepackage{graphics}
\usepackage{multirow}
\usepackage[colorlinks,plainpages=false]{hyperref}

\begin{document}
\hyphenpenalty=6000
\tolerance=1000


\title{Supercritically charged objects and electron-positron pair creation}


\author{Cheng-Jun~Xia$^{1}$}
\email{cjxia@nit.zju.edu.cn}
\author{She-Sheng~Xue$^{2}$}
\email{xue@icra.it; shesheng.xue@gmail.com}
\author{Ren-Xin~Xu$^{3,4}$}
\email{r.x.xu@pku.edu.cn}
\author{Shan-Gui~Zhou$^{5,6,7,8}$}
\email{sgzhou@itp.ac.cn}

\affiliation{$^{1}${School of Information Science and Engineering, Zhejiang University Ningbo Institute of Technology, Ningbo 315100, China}
             \\$^{2}$ICRANet and Department of Physics, Sapienza University of Rome, Rome 00185, Italy
             \\$^{3}$School of Physics, Peking University, Beijing 100871, China
             \\$^{4}$Kavli Institute for Astronomy and Astrophysics, Peking University, Beijing 100871, China
             \\$^{5}$CAS Key Laboratory of Theoretical Physics, Institute of Theoretical Physics, Chinese Academy of Sciences, Beijing 100190, China
             \\$^{6}$University of Chinese Academy of Sciences, Beijing 100049, China
             \\$^{7}$Center of Theoretical Nuclear Physics, National Laboratory of Heavy Ion Accelerator, Lanzhou 730000, China
             \\$^{8}$Synergetic Innovation Center for Quantum Effects and Application, Hunan Normal University, Changsha 410081, China}

\date{\today}

\begin{abstract}
We investigate the stability and $e^+e^-$ pair creation of supercritically charged superheavy nuclei, $ud$QM nuggets, strangelets, and strangeon nuggets based on the Thomas-Fermi approximation. The model parameters are fixed by reproducing masses and charge properties of these supercritically charged objects reported in earlier publications. It is found that $ud$QM nuggets, strangelets, and strangeon nuggets may be more stable than ${}^{56}$Fe at the baryon number $A\gtrsim 315$, $5\times10^4$, and $1.2\times10^8$, respectively. For those stable against neutron emission, the most massive superheavy element has a baryon number $\sim$965, while $ud$QM nuggets, strangelets, and strangeon nuggets need to have baryon numbers larger than $39$, 433, and $2.7\times10^5$. The $e^+e^-$ pair creation will inevitably start for superheavy nuclei with charge numbers $Z\geq177$, for $ud$QM nuggets with $Z\geq163$, for strangelets with $Z\geq 192$, and for strangeon nuggets with $Z\geq 212$. A universal relation $Q/R_e = \left(m_e - \bar{\mu}_e\right)/\alpha$ is obtained at a given electron chemical potential $\bar{\mu}_e$, where $Q$ is the total charge and $R_e$ the radius of electron cloud. The maximum number of $Q$ without causing $e^+e^-$ pair creation is then fixed by taking $\bar{\mu}_e=-m_e$. For supercritically charged objects with $\bar{\mu}_e<-m_e$, the decay rate for $e^+e^-$ pair production is estimated based on the Jeffreys-Wentzel-Kramers-Brillouin (JWKB) approximation. It is found that most positrons are emitted at $t\lesssim 10^{-15}$ s, while a long lasting positron emission can be observed for large objects with $R\gtrsim 1000$ fm. The emission of positrons and electron-positron annihilation from supercritically charged objects may be partially responsible for the short $\gamma$-ray burst during the merger of binary compact stars, the 511 keV continuum emission, as well as the narrow faint emission lines in X-ray spectra from galaxies and galaxy clusters.
\end{abstract}


\maketitle


\section{\label{sec:intro}Introduction}
The possible existence of objects heavier than the currently known nuclei has been a long-standing and intriguing question. As early as in 1960s, it was suggested that there may exist unusually stable or long-lived superheavy nuclei due to quantum shell effects, i.e., the island of stability of superheavy nuclei~\cite{Myers1966_NP81-1, Sobiczewski1966_PL22-500, Meldner1967_ArkivF36-593}. Based on cold and hot fusion reactions, superheavy elements with charge number $Z$ up to 118 have been synthesized~\cite{Hofmann2000_RMP72-733, Morita2004_JPSJ73-2593, Oganessian2007_JPG34-R165, Oganessian2010_PRL104-142502}. The quest to obtain heavier elements is still ongoing, which is focused both on their properties~\cite{Lalazissis1996_NPA608-202, Rutz1997_PRC56-238, Long2002_PRC65-047306, Ren2003_PRC67-064302, Zhang2005_NPA753-106, ZHANG2006_IJMPE15-1601, Meng2019_SCPMA63-212011} and synthesis mechanism~\cite{Li2003_EPL64-750, Zhang2006_PRC74-017304, Feng2007_PRC76-044606, SHEN2008_IJMPE17-66, Liu2009_PRC80-054608, Zagrebaev2010_NPA834-366c, Oganessian2009_PRC79-024603, Gates2011_PRC83-054618, Wang2012_PRC85-041601, Duellmann2013, Adamian2018_NPA970-22}. Meanwhile, there exist other possibilities. For example, it was argued that strange quark matter (SQM) comprised of approximately equal numbers of $u$, $d$, and $s$ quarks may be more stable than nuclear matter (NM)~\cite{Bodmer1971_PRD4-1601, Witten1984_PRD30-272, Terazaw1989_JPSJ58-3555}. This indicates the possible existence of stable SQM objects such as strangelets~\cite{Farhi1984_PRD30-2379, Berger1987_PRC35-213, Gilson1993_PRL71-332, Peng2006_PLB633-314},  nuclearites~\cite{Rujula1984_Nature312-734, Lowder1991_NPB24-177}, meteorlike compact ultradense objects~\cite{Rafelski2013_PRL110-111102}, and strange stars~\cite{Itoh1970_PTP44-291, Alcock1986_ApJ310-261, Haensel1986_AA160-121}. Nevertheless, if we consider the dynamical chiral symmetry breaking~\cite{Buballa1999_PLB457-261, Klahn2015_ApJ810-134}, the stability window of SQM vanishes. An interesting proposition was raised recently suggesting that quark matter comprised of only $u$ and $d$ quarks ($ud$QM) may be more stable~\cite{Holdom2018_PRL120-222001}. It was shown that the energy per baryon of $ud$QM nuggets become smaller than 930 MeV at $A\gtrsim 300$~\cite{Holdom2018_PRL120-222001}, while the properties of nonstrange quark stars are still consistent with current pulsar observations~\cite{Zhao2019_PRD100-043018, Zhang2020_PRD101-043003}. Inspired by various astrophysical observations~\cite{Lai2017_JPCS861-012027}, instead of deconfined quark matter, it was proposed that a solid state comprised of strangeons (quark-clusters with three-light-flavor symmetry) can be the true ground state~\cite{Xu2003_ApJ596-L59, Xu2018_SCPMA61-109531}, then small strangeon nuggets could also be stable and persist in the universe~\cite{Xu2019_AIPCP2127-020014}.

To synthesize these heavy objects with terrestrial experiments is very difficult. The fusion evaporation-residue cross sections in producing superheavy elements with $Z>118$ are extremely small and synthesizing them requires great efforts~\cite{Gates2011_PRC83-054618, Wang2012_PRC85-041601, Duellmann2013, Adamian2018_NPA970-22}. The possible production of strangelets via heavy-ion collisions was proposed in the 1980s~\cite{Greiner1987_PRL58-1825, Greiner1991_PRD44-3517}, while up till now no evidence of their existence is obtained~\cite{Abelev2007_PRC76-011901, Ellis2008_JPG35-115004}. Meanwhile, the $ud$QM nuggets and strangeon nuggets have not been observed in any of the heavy-ion collision experiments either. The situation may be very different in astrophysical environments. Being one type of the most dense celestial objects in the universe, pulsars provide natural laboratories for strongly interacting matter (termed simply strong matter) at the highest densities. As discussed in numerous investigations, pulsars are often recognized as neutron stars comprised of nuclear matter. Due to a first-order liquid-gas phase transition at subsaturation densities, nuclear matter could form pasta phase in the inner crust region of a neutron star~\cite{Ravenhall1983_PRL50-2066, Horowitz2005_PRC72-035801, Maruyama2005_PRC72-015802}, where giant nuclei with $Z$ up to $10^3$ are expected~\cite{Shen2011_ApJ197-20, Togashi2017_NPA961-78}. Meanwhile, if any of the arguments on SQM, $ud$QM, or strangeon matter (SM) is true, pulsars may in fact be strange stars~\cite{Itoh1970_PTP44-291, Alcock1986_ApJ310-261, Haensel1986_AA160-121, Weber2005_PPNP54-193, Jaikumar2006_PRL96-041101, Perez-Garcia2010_PRL105-141101, Herzog2011_PRD84-083002, Dexheimer2013_EPJC73-2569, Xu2015_PRD92-025025}, nonstrange quark stars~\cite{Zhao2019_PRD100-043018, Zhang2020_PRD101-043003}, or strangeon stars~\cite{Xu2003_ApJ596-L59, Lai2017_JPCS861-012027, Xu2018_SCPMA61-109531}.

The matter inside compact stars can be released during the merger of a binary system by both tidal disruption and squeezing as the stars come into contact~\cite{Baiotti2019_PPNP109-103714, Kasen2017_Nature551-80}. With a simple estimation on the balance between the tidal force and surface tension $\sigma$, the mass of the heaviest objects ejected into space is $M_\mathrm{max}\approx {3 R_\mathrm{c}^3\sigma}/{GM_\mathrm{c}}$, where $R_\mathrm{c}$ is the distance to the centre and $M_\mathrm{c}$ the total mass of the binary system. Nevertheless, in such a violent environment, the ejecta is heated and further collisions between those objects are expected, then most of the heavy objects are expected to decay. For example, in the binary neutron star merger event GW170817~\cite{LVC2017_PRL119-161101}, the ejecta quickly evolves into a standard neutron-rich environment for r-process nucleosynthesis and produces the transient counterpart AT2017gfo~\cite{Kasen2017_Nature551-80, Kasliwal2017_Science358-1559}, which is recently confirmed by the identification of the neutron-capture element strontium~\cite{Watson2019_Nature574-497}. For the merger of strange stars, strangelets are ejected but quickly evaporate into nucleons due to neutrino heating~\cite{Bucciantini2019}. Strangeon nuggets are formed during the merger of binary strangeon stars, and their decay provides an important energy source for the bolometric light curve of the following strangeon kilonova~\cite{Lai2018_RAA18-024}.

In such cases, even if heavy objects are ejected from compact stars, they may not survive since most of them decay into neutrons. However, if the charge number of those objects is large enough, a supercritical electric field can be built around them and lead to $e^+e^-$ pair production via the Schwinger mechanism~\cite{Schwinger1951_PR82-664}. During the merger of a binary system, large amount of matter ($\sim$$10^{-5}\text{--}10^{-2}\ M_\odot$) are ejected into space within a few seconds~\cite{Baiotti2019_PPNP109-103714, Kasen2017_Nature551-80}. Objects with various sizes are then formed in the ejecta, which will collide with each other and are usually heated. In such a catastrophic event, the electrons of those objects may be stripped away, which involve various possible mechanisms. For example: 1. The thermal ionization process should take effect at a high temperature~\cite{Usov1998_PRL80-230}; 2. When those objects cross areas with strong magnetic fields\footnote{The minimum magnetic field strength to create
supercritically charged objects in this scenario is roughly $3.4\times10^{12}$ G, which is obtained by equating the the Coulomb and Lorentz forces with the objects moving in a typical speed of 0.1$c$~\cite{Kasen2017_Nature551-80}.}, electrons are trapped along the magnetic field lines while the massive core passes through, i.e., the Lorentz ionization~\cite{Popov1997_PLA229-306}; 3. The collision with other objects, charged particles, and photons could excite the bound electrons into the continuum of free electron states~\cite{Post1977_ADNDT20-397, Mazzotta1998_AASS133-403}; 4. The Goldreich-Julian effect of electric charge separation should also play a role if the central merger remnant does not collapse promptly into a black hole~\cite{Goldreich1969_ApJ157-869}. In such cases, the charge number of those objects may increase significantly and exceed the critical values for $e^+e^-$ pair creation. Depending on the time of their creation, the emitted positrons may produce a distinct photon signature via positronium decay~\cite{Weidenspointner2006_AA450-1013, Prantzos2011_RMP83-1001}, or form an electron-positron plasma. Meanwhile, due to back-reaction the $e^+e^-$ pairs may create alternating electric fields in time, which emit electromagnetic radiations with the peak frequency located around 4 keV~\cite{Han2010_PLB691-99}. The corresponding signals for the existence of heavy objects may be identified based on various astrophysical observations. For the gravitational-wave event GW170817, a short $\gamma$-ray burst GRB 170817A that lasted about 2 s was observed shortly after ($1.74\pm0.05$ s), with a photon peak energy around 220 keV~\cite{Goldstein2017_ApJ848-L14, Abbott2017_ApJ848-L13, Rueda2018_JCAP2018-006}. In this work, we thus investigate the maximum charge numbers and the $e^+e^-$ pair creations for those heavy objects. The paper is organized as follows. In Sec.~\ref{sec:DM}, we present our theoretical framework to model the properties of NM, SQM, $ud$QM, and SM around their energy minimum. The properties of finite nuclei, $ud$QM nuggets, strangelets, and strangeon nuggets are then obtained in Sec.~\ref{sec:nuggets} based on the method adopted in our previous publications~\cite{Xia2016_SciBull61-172, Xia2016_SciSinPMA46-012021_E, Xia2016_PRD93-085025, Xia2017_NPB916-669}, and the $e^+e^-$ pair creations for supercritically charged objects are investigated in Sec.~\ref{sec:epem}. Our conclusion is given in Sec.~\ref{sec:con}.

\section{\label{sec:DM}Properties of strong matter}
The properties of various types of strong matter forming the supercritically charged objects can be well approximated by expanding the energy per baryon to the second order, i.e.,
\begin{equation}
\frac{E_\mathrm{DM}}{n_\mathrm{b}} = \varepsilon_0 + \frac{K_0}{18}\left(\frac{n_\mathrm{b}}{n_0}-1\right)^2 + 4\varepsilon_\mathrm{s} \left(f_Z - f_{Z0}\right)^2.  \label{eq:E_DM}
\end{equation}
Here $E_\mathrm{DM}$ is the energy density, $n_\mathrm{b}$ the baryon number density, and $f_Z$ the charge fraction with the charge density $f_Z n_\mathrm{b}$. The parameter $\varepsilon_0$ is the minimum energy per baryon at saturation density $n_0$ and charge fraction $f_{Z0}$, while $K_0$ is the incompressibility parameter and $\varepsilon_\mathrm{s}$ the symmetry energy. The exact values for those parameters are fixed according to the properties of strong matter obtained based on various studies. Note that Eq.~(\ref{eq:E_DM}) does not involve any information on the particles that the strong matter is made of, where the evolution of their masses and coupling constants are not explicitly shown. To obtain those properties, one should refer to the models that determine the parameters of Eq.~(\ref{eq:E_DM}). In this work, we adopt four representative parameter sets for NM, $ud$QM, SQM, and SM, which are summarized in Table~\ref{table:param}.

The baryon chemical potential $\mu_\mathrm{b} = \frac{\partial E_\mathrm{DM}}{\partial n_\mathrm{b}}$ and charge chemical potential $\mu_Q = \frac{1}{n_\mathrm{b}}\frac{\partial E_\mathrm{DM}}{\partial f_Z}$ of strong matter are obtained with
\begin{eqnarray}
\mu_\mathrm{b} &=&  \varepsilon_0 + \frac{K_0}{18}\left(3\frac{n_\mathrm{b}^2}{n_0^2} - 4\frac{n_\mathrm{b}}{n_0} +1\right)
+ 4\varepsilon_\mathrm{s} \left(f_{Z0}^2-f_Z^2\right), \label{eq:mub_DM}\\
\mu_Q &=& 8 \varepsilon_\mathrm{s} \left(f_Z-f_{Z0}\right). \label{eq:mue_DM}
\end{eqnarray}
Then the pressure is fixed according to the basic thermodynamic relations, i.e.,
\begin{equation}
P_\mathrm{DM} =\mu_\mathrm{b}n_\mathrm{b}+ \mu_Q f_Z n_\mathrm{b} - E_\mathrm{DM}=\frac{K_0 n_\mathrm{b}^2}{9 n_0^2} \left({n_\mathrm{b}}-{n_0}\right). \label{eq:P_DM}
\end{equation}

In nuclear matter, the minimum energy per baryon is obtained at $f_Z=f_{Z0} = 0.5$ and the saturation density $n_0\approx 0.15\text{--}0.16\ \mathrm{fm}^{-3}$, where $\varepsilon_0 = m_N - B$ with the binding energy $B\approx 16$ MeV, the incompressibility $K_0 = 240 \pm 20$ MeV~\cite{Shlomo2006_EPJA30-23}, and the symmetry energy $\varepsilon_\mathrm{s} = 31.7 \pm 3.2$ MeV~\cite{Li2013_PLB727-276, Oertel2017_RMP89-015007} are constrained with terrestrial experiments and nuclear theories. In this work, we take their central values with $n_0= 0.16\ \mathrm{fm}^{-3}$, $\varepsilon_0 = 922$ MeV, $K_0 = 240$ MeV, and $\varepsilon_\mathrm{s} = 31.7$ MeV.

The properties of $ud$QM obtained with linear sigma model in Ref.~\cite{Holdom2018_PRL120-222001} can be well reproduced if we take $n_0= 0.22\ \mathrm{fm}^{-3}$, $\varepsilon_0 = 887$ MeV, $K_0 = 2500$ MeV, and $\varepsilon_\mathrm{s} = 17.35$ MeV with $f_{Z0} = 0.5$. Note that the symmetry energy $\varepsilon_\mathrm{s}$ adopted here is small and contains only the kinetic term. In fact, extensive investigations on the values of $\varepsilon_\mathrm{s}$ were carried out in the past few years, e.g., those in Refs.~\cite{Chu2014_ApJ780-135, Jeong2016_NPA945-21, Chen2017_NPR34-20, Chu2019_PRC99-035802, Wu2019_AIPCP2127-020032}, where one may find a different value for $\varepsilon_\mathrm{s}$.

To fix the properties of SQM, we adopt the pQCD thermodynamic potential density with non-perturbative corrections~\cite{Xia2017_NPB916-669}, i.e.,
\begin{equation}
\Omega =  \Omega^\mathrm{pt} + B, \label{eq:omega_bg}
\end{equation}
where $\Omega^\mathrm{pt}$ is the pQCD thermodynamic potential density up to the order of $\alpha_\mathrm{s}$ in the $\overline{\mathrm{MS}}$ scheme~\cite{Fraga2005_PRD71-105014}.
The scale dependence of the strong coupling constant and quark masses is given by
\begin{eqnarray}
\alpha_\mathrm{s}(\bar{\Lambda})
  &=& \frac{1}{\beta_0 L}   \left(1- \frac{\beta_1\ln{L}}{\beta_0^2 L}\right),
\label{eq:alpha} \\
m_i(\bar{\Lambda})
  &=& \hat{m}_i \alpha_\mathrm{s}^{\frac{\gamma_0}{\beta_0}}
      \left[ 1 + \left(\frac{\gamma_1}{\beta_0}-\frac{\beta_1\gamma_0}{\beta_0^2}\right) \alpha_\mathrm{s} \right],
\label{eq:mi}
\end{eqnarray}
where $\beta_0={9}/{4\pi}$ and $\beta_1={4}/{\pi^2}$ for the $\beta$-function, $\gamma_0=1/\pi$ and $\gamma_1={91}/{24\pi^2}$ for the $\gamma$-function, and $L=2 \ln\left( \frac{\bar{\Lambda}}{\Lambda_{\overline{\mathrm{MS}}}}\right)$ with $\Lambda_{\overline{\mathrm{MS}}}$ being the $\overline{\mathrm{MS}}$ renormalization point.
The renormalization scale $\bar{\Lambda}$ is expanded with respect to the average value of quark chemical potentials. Its value to the first order is
\begin{equation}
  \bar{\Lambda} = C_0 + \frac{C_1}{3} \left(\mu_u + \mu_d + \mu_s\right). \label{eq:Lambda}
\end{equation}
In this work we take $C_0=1$ GeV, $C_1=4$, and $B^{1/4} = 138$ MeV, so that the most massive strange star can reach a mass of 2$M_\odot$~\cite{Xia2017_NPB916-669}. The parameters in Eq.~(\ref{eq:E_DM}) are then obtained by varying $\mu_u$ and $\mu_d$ ($\mu_d = \mu_s$) around the minimum energy per baryon, which is fixed at zero external pressure $P=-\Omega=0$ and chemical equilibrium $\mu_u = \mu_d = \mu_s$ for infinite strange quark matter. The Coulomb interaction is neglected here, which will be considered for finite sized objects. This gives $n_0= 0.296\ \mathrm{fm}^{-3}$, $\varepsilon_0 = 924.9$ MeV, $K_0 = 2266$ MeV, and $\varepsilon_\mathrm{s} = 18.2$ MeV with $f_{Z0} = 0.1$.

For strangeon matter, as was done in Ref.~\cite{Lai2009_MMRAS398-L31}, the potential energy density is obtained by adopting the Lennard-Jones potential between strangeons and assuming they form a simple-cubic structure. The energy density of strangeon matter is then obtained with
\begin{equation}
  E_\mathrm{SM} = 2 U_0 \left( 6.2 r_0^{12} n^5 - 8.4 r_0^6 n^3\right) + M_q n, \label{eq:E_strangeon}
\end{equation}
where $n = {n_\mathrm{b}}/{A_q}$ is the number density of strangeons. In this work we take the potential depth $U_0 = 50$ MeV, the range of interaction $r_0 = 2.63$ fm, the baryon number of a strangeon $A_q = 6$, and the mass of a strangeon $M_q = 975 A_q$ MeV. The obtained properties of strangeon stars well reproduce the current constraints on pulsar-like compact objects~\cite{Lai2019_EPJA55-60}. The energy density obtained with Eq.~(\ref{eq:E_strangeon}) around the saturation density can be approximated with Eq.~(\ref{eq:E_DM}) if we take $n_0= 0.27\ \mathrm{fm}^{-3}$, $\varepsilon_0 = 927.6$ MeV, and $K_0 = 4268$ MeV. Meanwhile, since stable strangeon matter is slightly positively charged due to the larger current mass of $s$-quarks, we take $f_{Z0} = 0.0063$ and $\varepsilon_\mathrm{s} = 250$ MeV.

\begin{table}
\caption{\label{table:param} The adopted parameter sets in Eq.~(\ref{eq:E_DM}) for nuclear matter (NM)~\cite{Li2013_PLB727-276, Oertel2017_RMP89-015007}, $ud$ quark matter ($ud$QM)~\cite{Holdom2018_PRL120-222001}, strange quark matter (SQM)~\cite{Xia2017_NPB916-669}, and strangeon matter (SM)~\cite{Lai2009_MMRAS398-L31}. }
\begin{tabular}{c|c|c|c|c|c|c} \hline \hline
         & $n_0$       & $f_{Z0}$ & $\varepsilon_0$ & $K_0$ & $\varepsilon_\mathrm{s}$ & $\sigma$  \\
         & fm${}^{-3}$ &          &      MeV        &   MeV &         MeV              &   MeV/fm${}^{2}$ \\   \hline
NM       &  0.16       &   0.5    &      922        &   240 &         31.7             &   1.34  \\
$ud$QM   &  0.22       &   0.5    &      887        &  2500 &        17.35             &   19.35 \\
SQM      &  0.296      &   0.1    &     924.9       &  2266 &        18.2              &   15  \\
SM       &  0.27       & 0.0063   &     927.6       &  4268 &         250              &   100  \\
\hline
\end{tabular}
\end{table}

Since the strong matter considered here is positively charged with $f_{Z0}>0$, the contribution of electrons should be considered due to the attractive Coulomb interaction. The electron energy density is obtained with
\begin{eqnarray}
E_e &=& \int_0^{\nu_e} \frac{p^2}{\pi^2} \sqrt{p^2+m_e^2}\mbox{d}p \nonumber \\
    &=&  \frac{m_e^4}{8\pi^{2}} \left[x_e(2x_e^2+1)\sqrt{x_e^2+1}-\mathrm{arcsh}(x_e) \right]. \label{eq:E_e}
\end{eqnarray}
Here $x_e\equiv \nu_e/m_e$ with $\nu_e$ being the Fermi momentum of electrons and $m_e=0.511$ MeV the electron mass. The number density, chemical potential, and pressure of electron gas are given by
\begin{eqnarray}
n_e &=& \nu_e^3/3\pi^2, \label{eq:ne}\\
\mu_e &=& \sqrt{\nu_e^2+m_e^2}, \label{eq:mue}\\
P_e &=& \mu_e n_e - E_e. \label{eq:Pe}
\end{eqnarray}
To reach the energy minimum, electrons interact with strong matter and the $\beta$-stability condition should be fulfilled, i.e.,
\begin{equation}
\mu_e = -\mu_Q. \label{eq:bstable_SM}
\end{equation}

\section{\label{sec:nuggets}Finite-sized objects}

To investigate the properties of finite-sized objects, we assume they are spherically symmetric and each of them consists of a core of strong matter surrounded by an electron cloud.
We thus adopt a unified description that was previously intended for SQM objects, i.e., the UDS model~\cite{Xia2016_SciBull61-172, Xia2016_SciSinPMA46-012021_E, Xia2016_PRD93-085025, Xia2017_NPB916-669}. The mass $M$, total baryon number $A$, net charge number $Z$, total charge number $Q$, and electron number $N_e$ of the object are determined by
\begin{eqnarray}
M &=& \int_0^\infty \left[4\pi r^2 E(r)+ \frac{r^2}{2 \alpha}\left(\frac{\mbox{d} \varphi}{\mbox{d}r}\right)^2 \right] \mbox{d}r + 4\pi R^2 \sigma, \label{eq:mass}\\
A &=& \int_0^R 4\pi r^2  n_\mathrm{b}(r) \mbox{d}r, \label{eq:A}\\
Z &=& \int_0^R 4\pi r^2 f_Z(r) n_\mathrm{b}(r) \mbox{d}r, \label{eq:Z}\\
Q &=& \int_0^\infty 4\pi r^2  n_\mathrm{ch}(r) \mbox{d}r, \label{eq:Q}\\
N_e &=& \int_0^\infty 4\pi r^2  n_e(r) \mbox{d}r = Z - Q. \label{eq:Ne}
\end{eqnarray}
Note that the local energy density is obtained with $E = E_\mathrm{DM}+E_e$ and charge density $n_\mathrm{ch} = f_Z n_\mathrm{b} - n_e$ at $r\leq R$, while the region at $r>R$ is occupied by electrons with $E = E_e$ and $n_\mathrm{ch} = - n_e$. The energy densities for strong matter $E_\mathrm{DM}$ and electrons $E_e$ are obtained with Eqs.~(\ref{eq:E_DM}) and (\ref{eq:E_e}), while the electron density is determined by Eq.~(\ref{eq:ne}). The finite-size effects are treated with a surface tension $\sigma$, which accounts for the energy contribution from density gradient terms of strong interaction. By minimizing the mass in Eq.~(\ref{eq:mass}) based on the Thomas-Fermi approximation, we obtain the density distributions $n_\mathrm{b}(r)$,  $f_Z(r) n_\mathrm{b}(r)$, and $n_e(r)$ ($\mu_e = -\mu_Q$), which follows
\begin{eqnarray}
&& \mu_\mathrm{b}(r) = \mathrm{constant}, \label{eq:mubdis} \\
&& \bar{\mu}_e = \mu_e(r) - \varphi(r) = \mathrm{constant},  \label{eq:pdis}
\end{eqnarray}
with the electric potential $\varphi(r)$ determined by
\begin{equation}
r^2  \frac{\mbox{d}^2 \varphi}{\mbox{d}r^2} + 2r\frac{\mbox{d} \varphi}{\mbox{d}r} + 4\pi\alpha r^2 n_\mathrm{ch} = 0. \label{eq:def0}
\end{equation}
Here $\mu_\mathrm{b}$ and $\bar{\mu}_e$ correspond to the respective chemical potentials of finite-sized objects. The charge density is obtained with $n_\mathrm{ch}(r) = f_Z(r) n_\mathrm{b}(r) - n_e(r)$. With the local chemical potentials determined by Eq.~(\ref{eq:pdis}), the local density profiles are then obtained based on the properties predicted in Sec.~\ref{sec:DM}. At a given surface tension value $\sigma$, the radius of the core $R$ is fixed according to the dynamic stability of the hadron/quark-vacuum interface, i.e.,
\begin{equation}
P_\mathrm{DM}(R) = \frac{2\sigma}{R}. \label{eq:P_stable}
\end{equation}
In our calculation, electrons are trapped within the Coulomb potential of the core and $\bar{\mu}_e$ represents the top of the Fermi sea for electrons. By increasing $\bar{\mu}_e$, the total number of electrons $N_e$ increases, which reduces the total charge number with $Q=Z-N_e$. The boundary of the electron cloud $R_e$ is fixed at vanishing $n_e$, i.e., $\mu_e(R_e) = m_e$. In fact, since there is no electron persists at $r>R_e$, the Coulomb potential is simply $\varphi(r)={\alpha Q}/{r}$. According to Eq.~(\ref{eq:pdis}), at given $\bar{\mu}_e$ one obtains the following relation
\begin{equation}
  \frac{Q}{R_e} = \frac{m_e - \bar{\mu}_e}{\alpha}. \label{eq:QR_relation}
\end{equation}
If the core radius exceeds the Bohr radius (e.g., $R\gtrsim 10^5$ fm), we have $R\approx R_e$ and a direct correlation between $Q$ and $\bar{\mu}_e$ can be obtained with
\begin{equation}
  Q =(m_e - \bar{\mu}_e) \frac{R}{\alpha}. \label{eq:QR_relationM}
\end{equation}

Based on the parameter sets indicated in Table~\ref{table:param}, we can study finite-sized objects comprised of NM, $ud$QM, SQM, and SM, i.e., finite nuclei, $ud$QM nuggets, strangelets, and strangeon nuggets. For finite nuclei, to reproduce the masses of known atomic nuclei~\cite{Audi2017_CPC41-030001, Huang2017_CPC41-030002, Wang2017_CPC41-030003}, we take $\sigma  = 1.34 \ \mathrm{MeV/fm}^{2}$. The surface tension value for $ud$QM nuggets is indicated in Ref.~\cite{Holdom2018_PRL120-222001} with $\sigma  = 19.35 \ \mathrm{MeV/fm}^{2}$. For strangelets, it was shown that the curvature term is important for small strangelets~\cite{Madsen1993_PRL70-391}. However, small strangelets are unstable according to our previous calculation~\cite{Xia2017_NPB916-669}, we thus neglect the curvature term and take $\sigma  = 15 \ \mathrm{MeV/fm}^{2}$, which well reproduces the strangelets' masses at $A\gtrsim 200$. The surface tension value $\sigma$ for strangeon nuggets is not determined and should be fixed based on the interaction between strangeons~\cite{Guo2014_CPC38-055101}. In this work, however, we take a reasonable surface tension value $\sigma = 100$ MeV/fm${}^2$ since strangeon matter is in a solid-state. The adopted surface tension values are summarized in Table~\ref{table:param}.

At given $\mu_\mathrm{b}$ and $\bar{\mu}_e$, Eq.~(\ref{eq:def0}) is solved numerically and the density profiles are obtained according to Eq.~(\ref{eq:pdis}). The properties of a finite-sized object is then fixed based on Eqs.~(\ref{eq:Q}-\ref{eq:P_stable}). It is found that varying $\bar{\mu}_e$ has little impact on the obtained masses of finite-sized objects. To investigate the properties of supercritically charged objects, we thus adopt $\bar{\mu}_e=-m_e$ in our calculation.

\begin{figure}
\includegraphics[width=\linewidth]{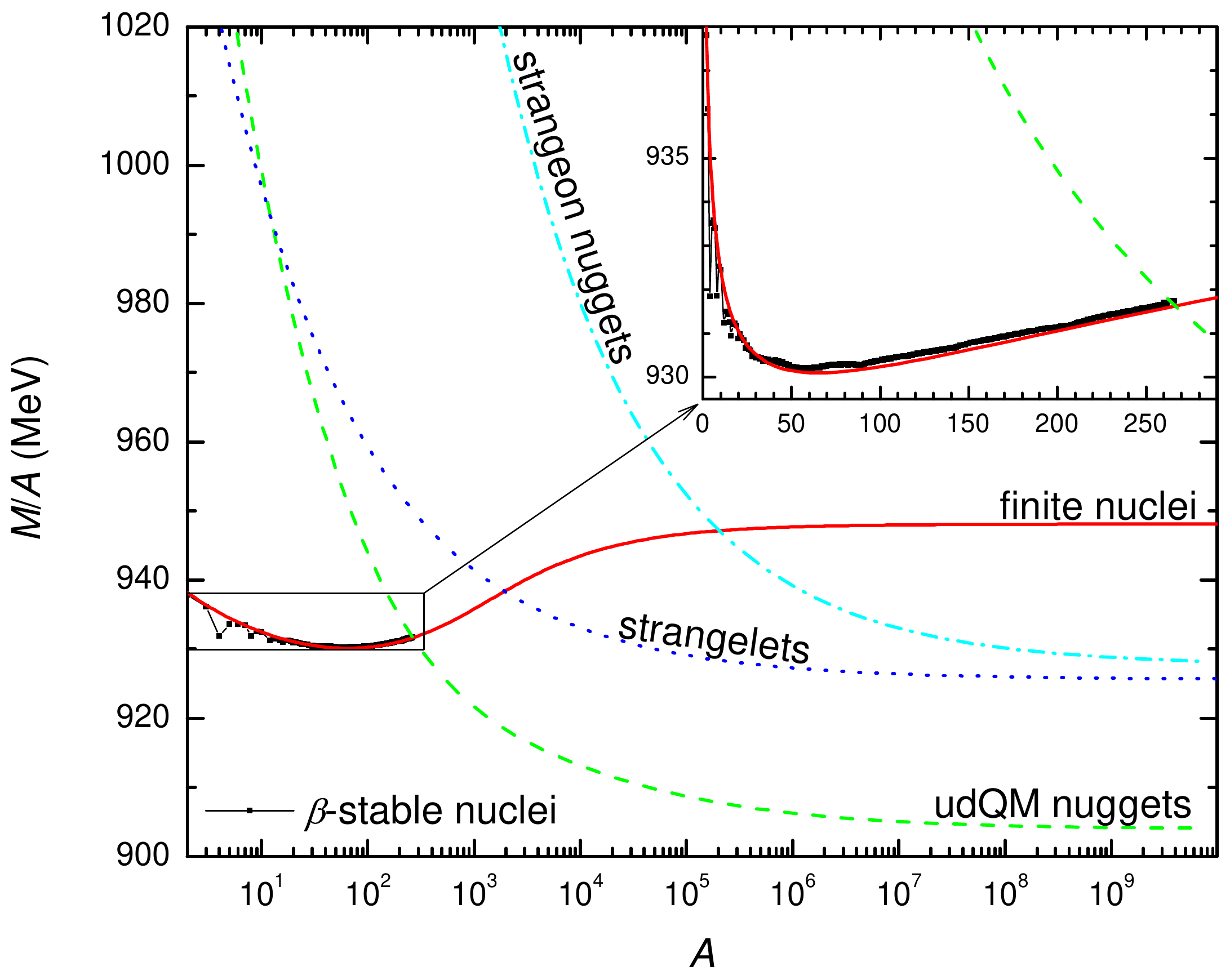}
\caption{\label{Fig:EpA} Energy per baryon for four types of finite-sized objects as functions of the baryon number $A$. The experimental data for $\beta$-stable nuclei are indicated with solid squares, which are obtained from the 2016 Atomic Mass Evaluation~\cite{Audi2017_CPC41-030001, Huang2017_CPC41-030002, Wang2017_CPC41-030003}.}
\end{figure}

In Fig.~\ref{Fig:EpA} we present the energy per baryon of finite-sized objects fulfilling the $\beta$-stability condition. The experimental values for finite nuclei obtained from the 2016 Atomic Mass Evaluation~\cite{Audi2017_CPC41-030001, Huang2017_CPC41-030002, Wang2017_CPC41-030003} are well reproduced in our framework. A minimum value corresponding to ${}^{56}$Fe is identified with $M/A=930$ MeV, which is mainly due to the small surface tension of nuclear matter. For other exotic objects such as $ud$QM nuggets, strangelets, and strangeon nuggets, the obtained energy per baryon is decreasing with $A$ due to the dominant surface energy correction. As indicated in Table~\ref{table:crit}, a critical baryon number $A_\mathrm{crit}$ can then be fixed for those objects, where at $A>A_\mathrm{crit}$ they become more stable than ${}^{56}$Fe, i.e., $M/A<930$ MeV. Note that the critical baryon number may vary with surface tension. In fact, if a small enough $\sigma$ is adopted, it was shown there also exists a local energy minimum for strangelets, where strangelets of a certain size are more stable than others~\cite{Alford2006_PRD73-114016, Jaikumar2006_PRL96-041101, Xia2017_JPCS861-012022}. Similar situations may occur for other exotic objects. A crossing between the curves of finite nuclei and $ud$QM nuggets is found at $A\approx 266$. In such cases, with the heaviest element ${}^{294}$Og synthesized by far~\cite{Oganessian2012_PRL109-162501}, producing $ud$QM nuggets may be imminent via heavy ion collisions or the decay of superheavy elements if $ud$QM is the true ground state for strong matter.

\begin{table}
\caption{\label{table:crit} The ranges of baryon ($A$) and/or charge ($Z$) numbers for objects that are stable against decaying into ${}^{56}$Fe with $M/A<930$ MeV, neutron emission with $S_n>0$, and $e^+e^-$ pair creation with $Z-Q<2$.}
\begin{tabular}{c|c|c|c|c} \hline \hline
                   &  $\frac{M}{A}<930$ &    $S_n>0$         &    \multicolumn{2}{c}{$Z-Q<2$}                  \\  \cline{2-5}
                   &           $A$      &    $A$             &     $Z$     &     $A$           \\  \hline
finite nuclei      &                    &  $< 965$           &   $<177$ & $\lesssim480$   \\
$ud$QM nuggets     &  $> 315$           &  $> 39$            &   $<163$ & $\lesssim609$   \\
strangelets        &  $> 5\times10^4$   &  $> 433$           &   $<192$ & $\lesssim16285$ \\
strangeon nuggets  &  $> 1.2\times10^8$ &  $> 2.7\times10^5$ &   $<212$ & $\lesssim90796$ \\
\hline
\end{tabular}
\end{table}

\begin{figure}
\includegraphics[width=\linewidth]{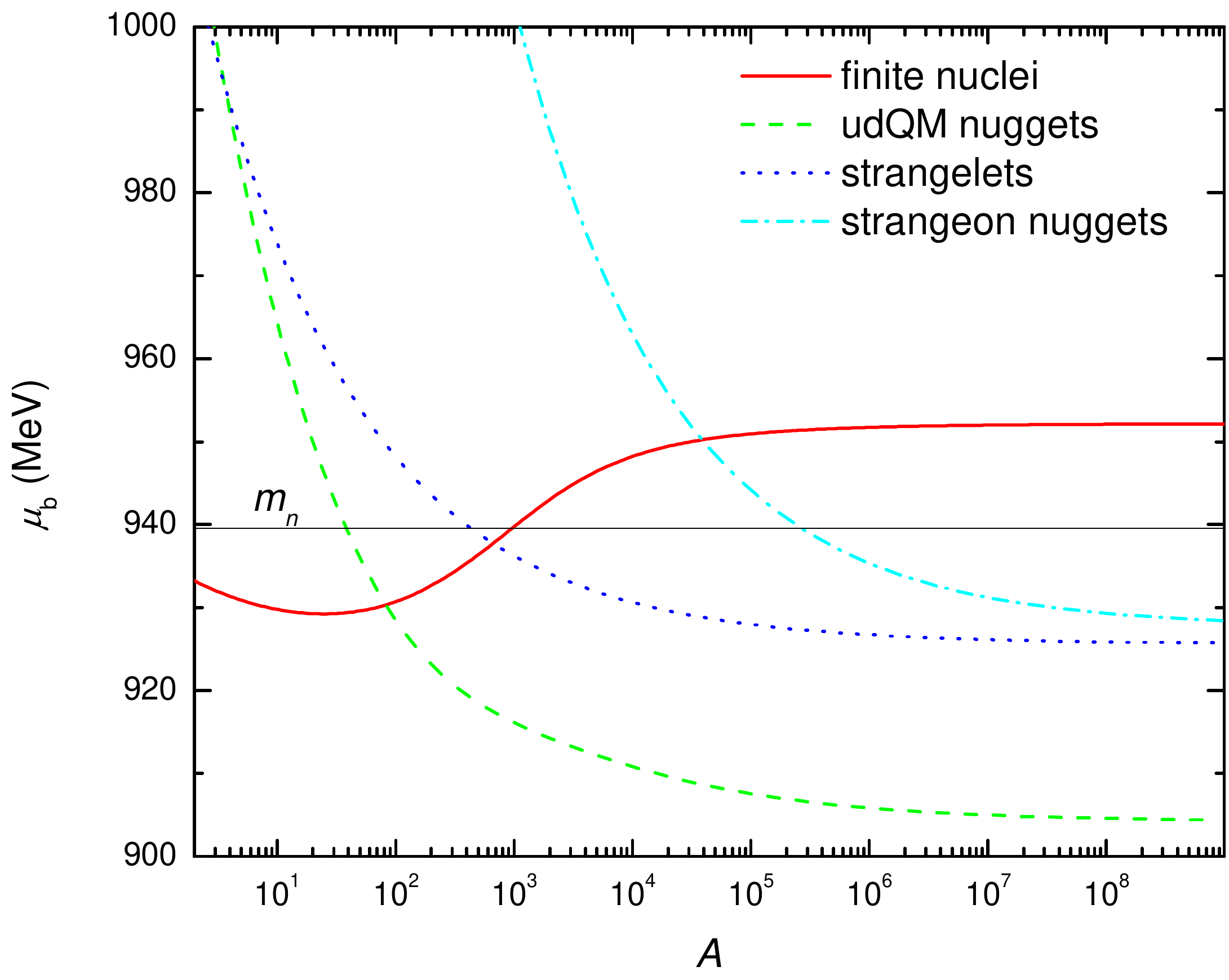}
\caption{\label{Fig:mub} Baryon chemical potential of finite-sized objects as functions of the baryon number $A$. The mass of a free neutron $m_n$ is indicated with the horizontal line.}
\end{figure}

The stability of those objects against particle emission can be observed through their chemical potentials. In Fig.~\ref{Fig:mub} we present the baryon chemical potential $\mu_\mathrm{b}$ as functions of the baryon number $A$. The neutron separation energy is then obtained with $S_n = m_n - \mu_\mathrm{b}$, which becomes negative once $\mu_\mathrm{b}>m_n$ and spontaneous neutron emission is thus inevitable for those objects once ejected into space. The corresponding baryon number ranges for objects that are stable against neutron emission ($S_n>0$) are listed in Table~\ref{table:crit}. For the emission of charged particles such as protons and $\alpha$ particles, the existence of a Coulomb barrier effectively reduces the rate of emission, which is less significant compared with neutron emissions at $S_n<0$. For superheavy elements with $A < 965$, however, the emission of charged particles as well as spontaneous fission should play important roles on their stability, which is expected to be sensitive to the shell effects and pairing. A more detailed investigation on these aspects is thus necessary, e.g., those in Refs.~\cite{Moller1976_PRL37-1461, Lalazissis1996_NPA608-202, Rutz1997_PRC56-238, Long2002_PRC65-047306, Ren2003_PRC67-064302, Zhang2005_NPA753-106, ZHANG2006_IJMPE15-1601, Xia2011_SCPMA54-109, Koura2014_PTEP2014-113D02}.

\begin{figure}
\includegraphics[width=\linewidth]{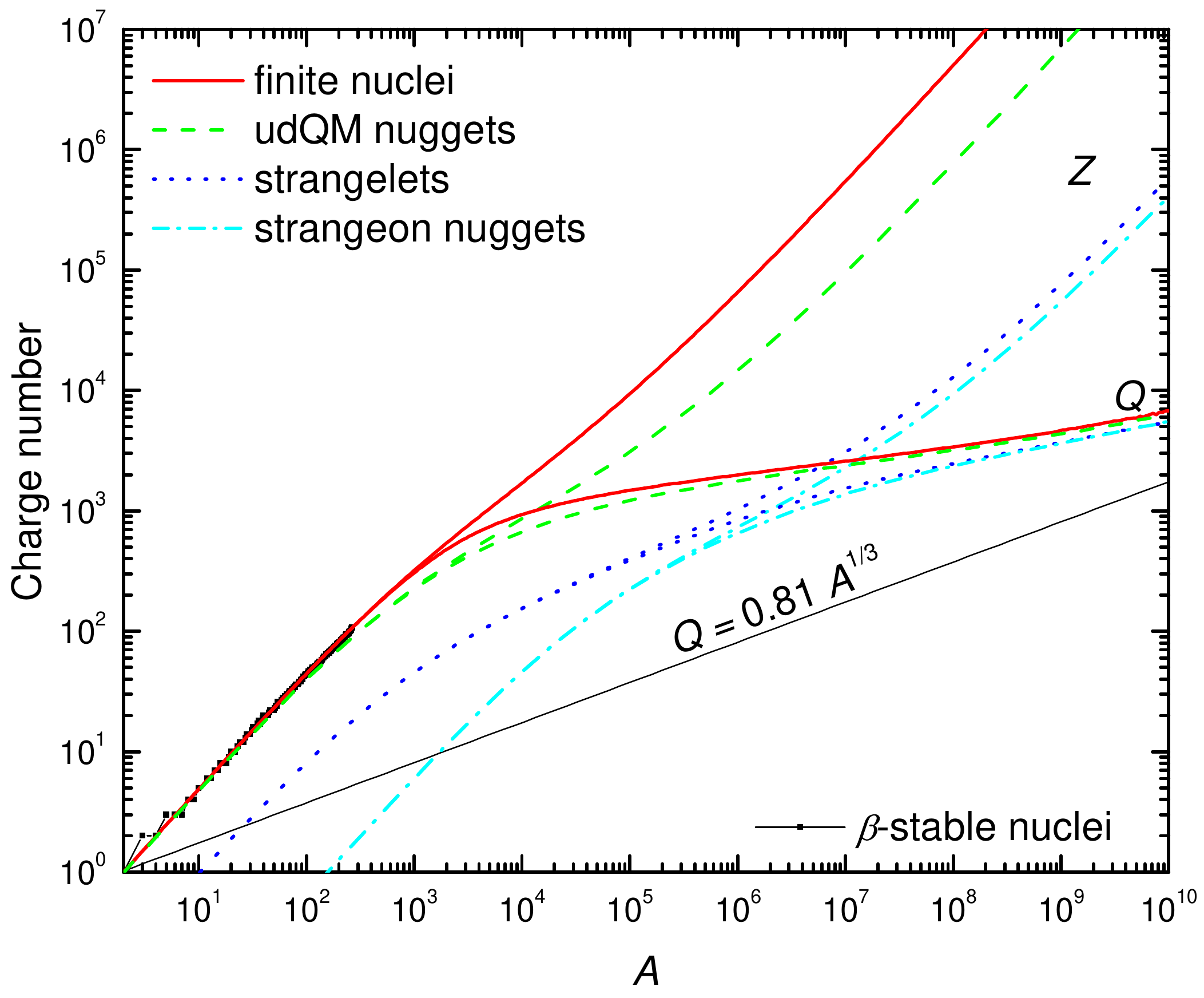}
\caption{\label{Fig:Charge} The net ($Z$) and maximum ($Q=Z-N_e$) charge numbers of finite-sized objects as functions of the baryon number $A$, obtained by taking $\bar{\mu}_e=-m_e$.}
\end{figure}

With a given electron potential $\bar{\mu}_e$, the structures of the core and electron cloud are obtained by solving Eq.~(\ref{eq:def0}). The net charge number $Z$ of the core is determined by subtracting the contributions of electrons, while the total charge number $Q$ includes contributions of all charged particles. As indicated in Eq.~(\ref{eq:QR_relation}), taking $\bar{\mu}_e=m_e$ neutralizes the core entirely and corresponds to the global charge neutrality condition with $Q=0$, while here we have adopted $\bar{\mu}_e=-m_e$, i.e., the upper edge of the electron Dirac sea. In such cases, $Q$ represents the maximum charge number without causing $e^+e^-$ pair creation. The obtained net and maximum charge numbers are presented in Fig.~\ref{Fig:Charge}. The predicted proton numbers for nuclei coincide with the experimental $\beta$-stability line as indicated with solid squares. For $ud$QM nuggets, the obtained charge numbers are slightly smaller than finite nuclei, which is mainly due to the small symmetry energy adopted here. By taking $f_{Z0} = 0.1$ and 0.0063 instead of 0.5, the obtained charge numbers for strangelets and strangeon nuggets are much smaller than the two-flavor cases. The net charge-to-mass ratios
vary smoothly from $f_{Z0}$ at $A\lesssim 100$ to small values at $A\gtrsim 10^9$, which are presented in Table~\ref{table:charge}. Meanwhile, as was discussed in our previous works~\cite{Xia2016_SciBull61-172, Xia2016_SciSinPMA46-012021_E, Xia2016_PRD93-085025, Xia2017_NPB916-669}, a constant surface charge density $Q(R)/R^2$ (as indicated in Table~\ref{table:charge}) is obtained at $A\gtrsim 10^9$ if we also consider the contribution of electrons in the core.

\begin{table}
\caption{\label{table:charge} The charge properties of maximum charged objects obtained at $\bar{\mu}_e=-m_e$, i.e., the net charge-to-mass ratios $Z/A$, the surface charge density of the core $Q(R)/R^2$ ($R$ in fm), and the ratio of maximum charge number to baryon number $Q/A^{1/3}$.}
\begin{tabular}{c|c|c|c|c} \hline \hline
                   &       \multicolumn{2}{c|}{$Z/A$}      &   $Q(R)/R^2$     &    $Q/A^{1/3}$           \\     \cline{2-5}
                   & $A\lesssim 100$ &   $A\gtrsim 10^9$   & $A\gtrsim 10^9$  &    $A\gtrsim 10^{15}$    \\    \hline
finite nuclei      &        0.5      &      0.047          &       1.4        &         0.81             \\
$ud$QM nuggets     &        0.5      &      0.0064         &      0.56        &         0.73             \\
strangelets        &        0.1      & $4.6\times 10^{-5}$ &      0.028       &         0.66             \\
strangeon nuggets  &      0.0063     & $3.2\times 10^{-5}$ &      0.020       &         0.68             \\
\hline
\end{tabular}
\end{table}

Since the single particle levels for electrons are degenerate in spin, a critical charge number $Z_\mathrm{crit}$ is obtained at $Z-Q=2$ according to Fig.~\ref{Fig:Charge}. The corresponding upper limits of baryon and charge numbers for objects that are stable against $e^+e^-$ pair creation with $Z-Q\leq2$ are presented in Table~\ref{table:crit}. For objects with larger $Z$, with the critical electric field built around the core, electrons will inevitably appear due to $e^+e^-$ pair creation, which effectively reduces the charge number from $Z$ to $Q$ ($Q<Z$). The corresponding decay rates can be estimated by Eq.~(\ref{eq:W_epm}). Note that the critical charge number for superheavy elements was a long-standing problem and many efforts were made in the past decades. For example, the critical charge number $Z_\mathrm{crit} = 137$ is obtained for a pointlike nucleus~\cite{Gordon1928_ZP48-11, Gordon1928_ZP48-180}. For more realistic cases, adopting different radii for finite-sized nuclei predicts various critical charge numbers with $Z_\mathrm{crit} = 171$--178~\cite{Ruffini2010_PR487-1, Kuleshov2015_PU58-785, Voronov2016_TMP187-633}, while our prediction in Fig.~\ref{Fig:Charge} with $Z_\mathrm{crit} = 177$ lies within this range. Finally, the maximum charge numbers $Q$ for different types of objects are converging at $A\gtrsim 10^8$ or $R\gtrsim 1000$ fm, where the variations on the core structures become insignificant.

\begin{figure}
\includegraphics[width=\linewidth]{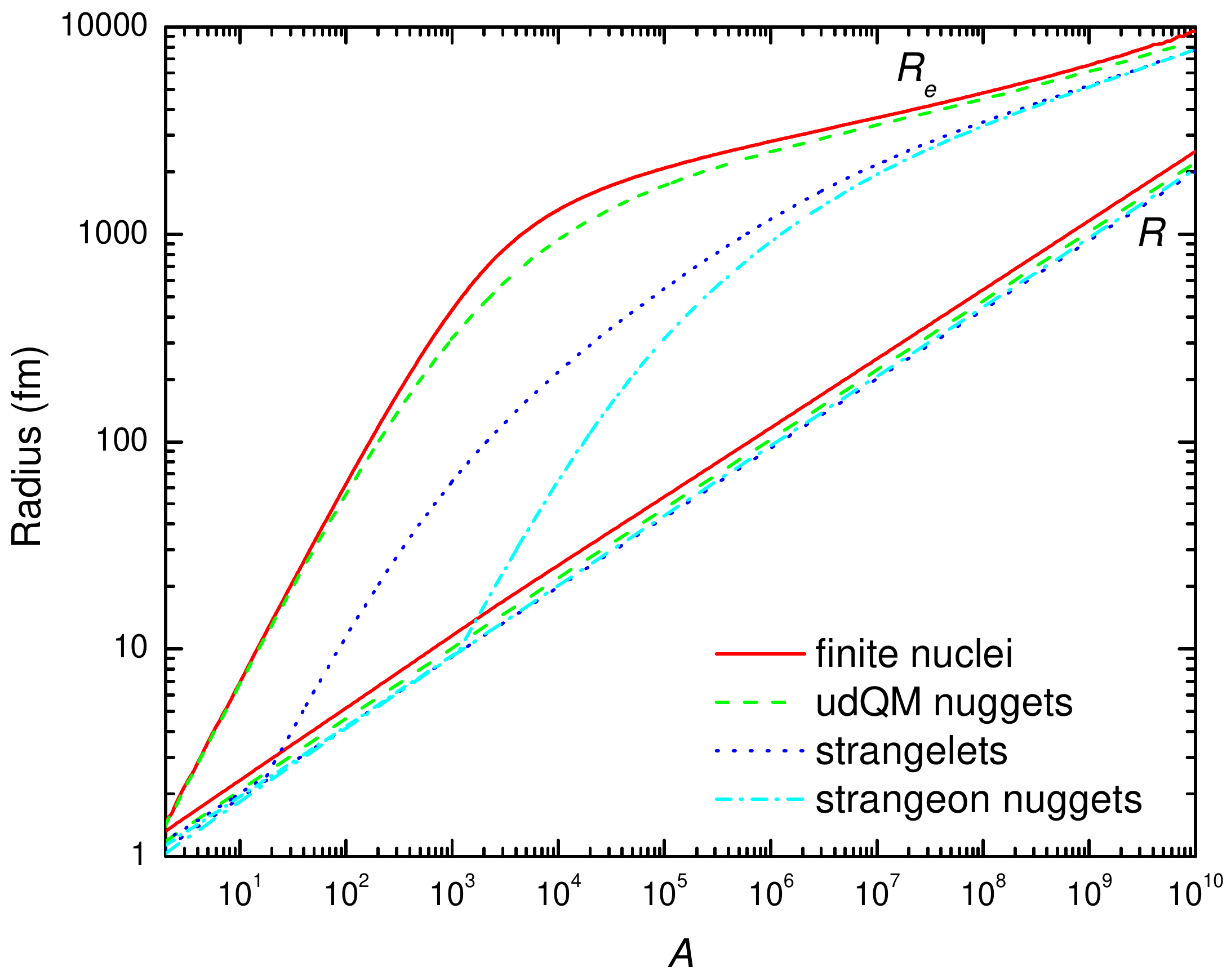}
\caption{\label{Fig:Radius} Radii of the core $R$ and electron cloud $R_e$ as functions of the baryon number $A$, obtained by taking $\bar{\mu}_e=-m_e$.}
\end{figure}

In Fig.~\ref{Fig:Radius} we present the radii of the core $R$ and electron cloud $R_e$ obtained at $\bar{\mu}_e=-m_e$. At $A\lesssim 4000$, the ratio of radius to baryon number $r_0=R/A^{1/3}$ is increasing with $A$, which arises from the Coulomb repulsion and a decrease of pressure from surface energy as in Eq.~(\ref{eq:P_stable}). In fact, such a decrease of baryon density was pointed out in previous studies, e.g., the bubble-like structures found in very heavy nuclei embedded in an electron background~\cite{Ebel2018_EPJA54-27}. Based on Eq.~(\ref{eq:QR_relation}), the maximum charge an object can carry without causing $e^+e^-$ pair production can then be obtained by taking $\bar{\mu}_e = - m_e$, which gives $Q = 0.71 R_e$ ($R_e$ in fm)~\cite{Madsen2008_PRL100-151102}. The radii of electron cloud $R_e$ are thus linked with the maximum charge number $Q$, which is indeed the case according to our numerical calculation. The relation also predicts the trend on the maximum charge numbers with $Q = 0.71 R$ (or $Q = 0.71 r_0 A^{1/3}$ with $r_0^3 \approx 3/4\pi n_0$) as we increase $A$, which should be valid at $R\gtrsim 10^5$ fm or $A\gtrsim 10^{15}$ with $R$ and $R_e$ being nearly the same. For finite nuclei, as indicated in Fig.~\ref{Fig:Charge}, adopting $n_0 = 0.16$ fm${}^{-3}$ gives $Q = 0.81 A^{1/3}$. For other exotic objects, as indicated in Table~\ref{table:charge}, $Q$ is smaller due to larger values for $n_0$.

\section{\label{sec:epem}$e^+e^-$ pair production}
\begin{figure}
\includegraphics[width=\linewidth]{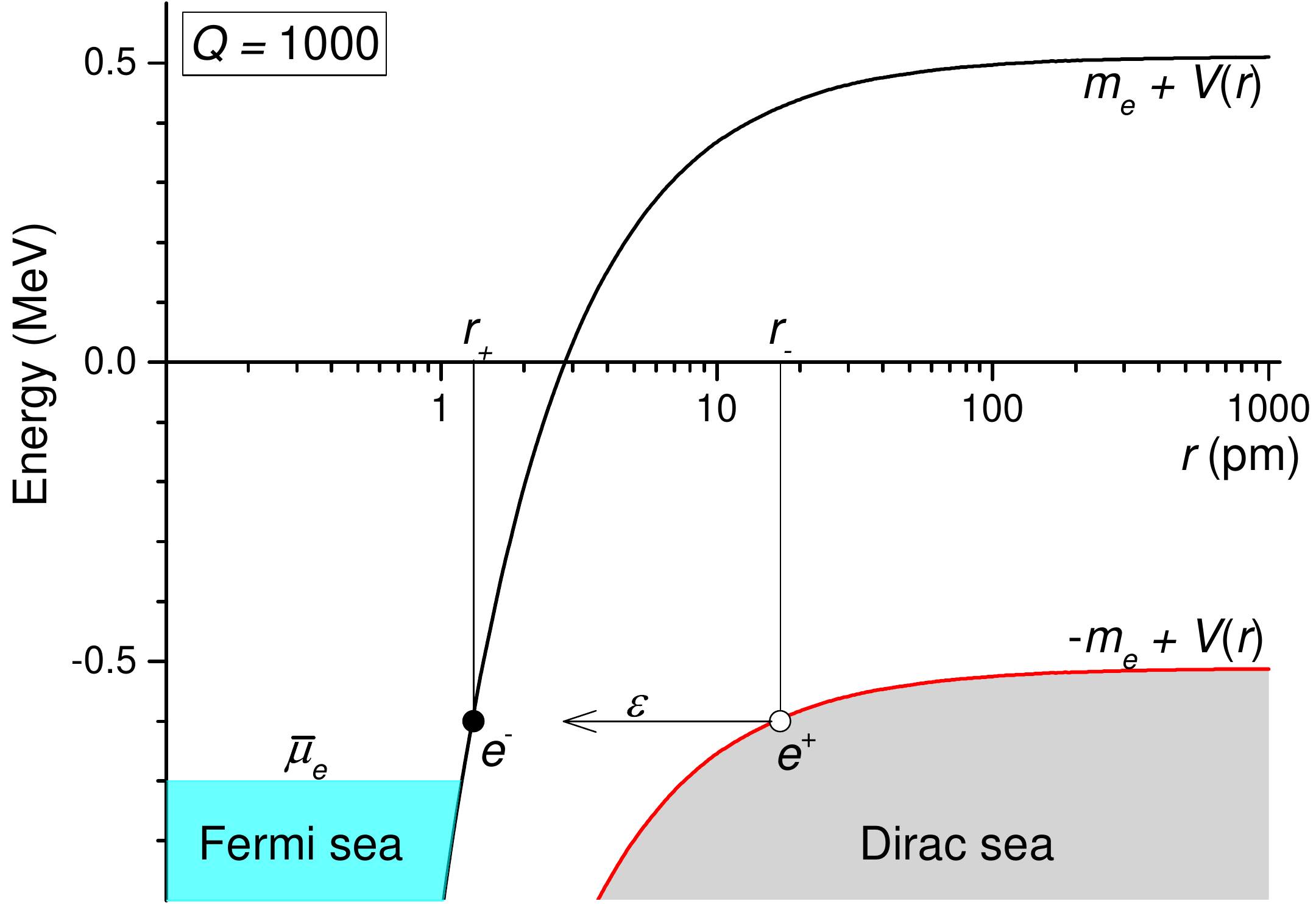}
\caption{\label{Fig:Schwinger} Positive and negative energy spectra for electrons in the Coulomb potential of a charged object with $Q=1000$.}
\end{figure}

For $e^+e^-$ pair production in the electric field of a positively charged object, an example of the tunneling process is illustrated in Fig.~\ref{Fig:Schwinger}.
Electrons located in the Dirac sea propagate into the Fermi sea (from $r_-$ to $r_+$), leaving behind a hole at $r_-$, i.e., positrons. The electron chemical potential of the system is $\bar{\mu}_e$ ($\leq -m_e$), with the total charge number $Q$. A potential for electrons is then obtained with $V(r)=-\varphi(r)=-{\alpha Q}/{r}$ for $r\ge R_e$. Note that the screening effects of electrons are included in the total charge number, where the charge number without electrons $Z$ is larger than $Q$. The tunneling process is only possible for electrons with energy $\bar{\mu}_e\leq \varepsilon \leq -m_e$, where the levels at $\varepsilon \leq \bar{\mu}_e$ are already occupied. According to the Thomas-Fermi approximation, a boundary for electrons is obtained at $r=R_e$ with $\bar{\mu}_e = V(R_e) + m_e$, beyond which electrons do not exist. The relation between $Q$, $R_e$, and $\bar{\mu}_e$ is indicated in Eq.~(\ref{eq:QR_relation}), while the maximum charge an object can carry without causing $e^+e^-$ pair production was obtained by taking $\bar{\mu}_e = - m_e$~\cite{Madsen2008_PRL100-151102}.

The decay rate of the vacuum for $e^+e^-$ pair production in an arbitrary constant electric field $\mathbf{E}$ is given by~\cite{Schwinger1951_PR82-664}
\begin{equation}
 \frac{\Gamma}{V} = \frac{\alpha \mathbf{E}^2}{\pi^2}\sum_{n=1}^\infty \frac{1}{n^2} \exp\left(-\frac{n\pi \mathbf{E}_c}{\mathbf{E}}\right),
\end{equation}
where the critical electric field is $\mathbf{E}_c=m_e^2/e = m_e^2/\sqrt{4\pi\alpha}$.

\begin{figure*}[htbp]
\begin{minipage}[t]{0.49\linewidth}
\centering
\includegraphics[width=\textwidth]{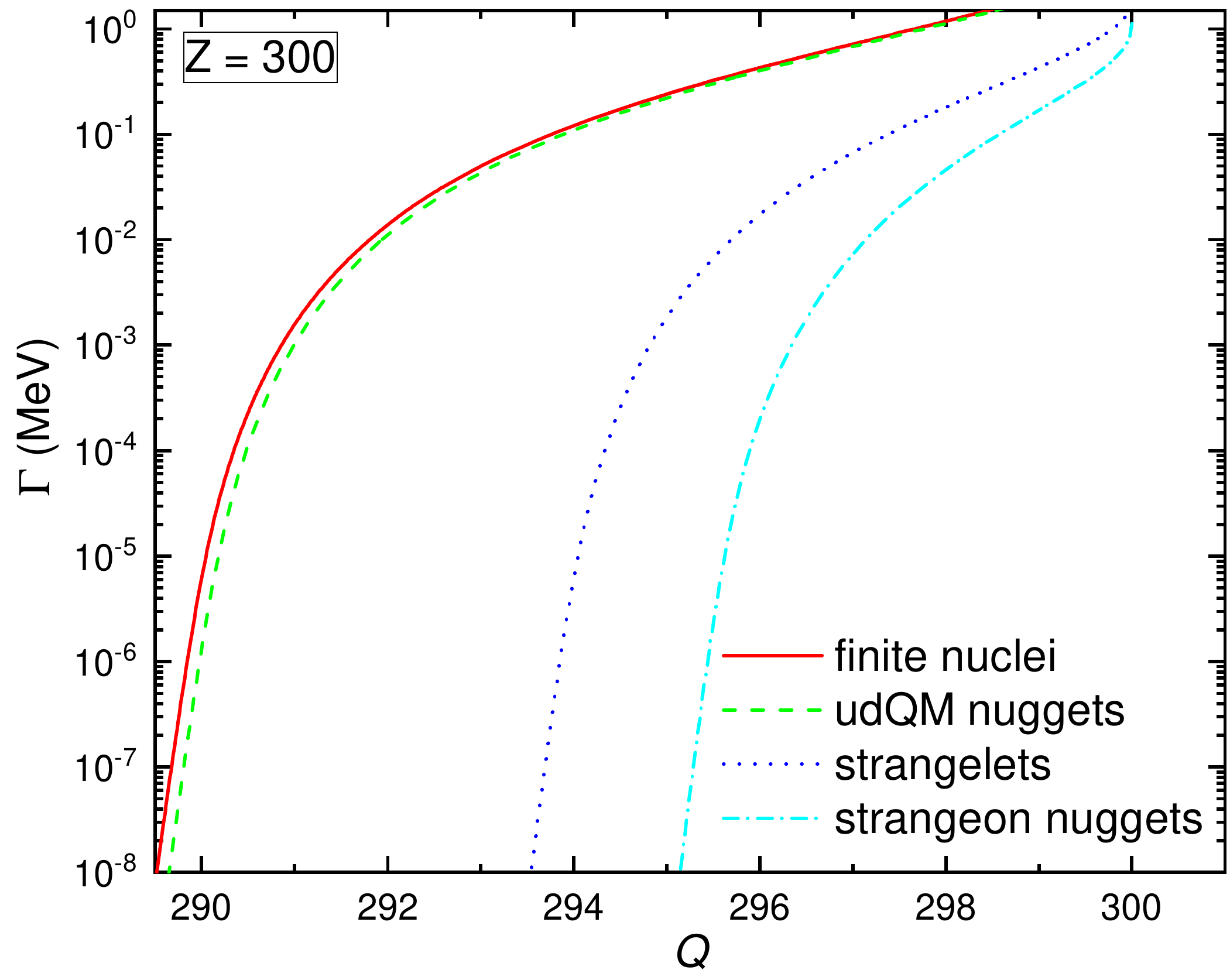}
\end{minipage}%
\hfill
\begin{minipage}[t]{0.51\linewidth}
\centering
\includegraphics[width=1\textwidth]{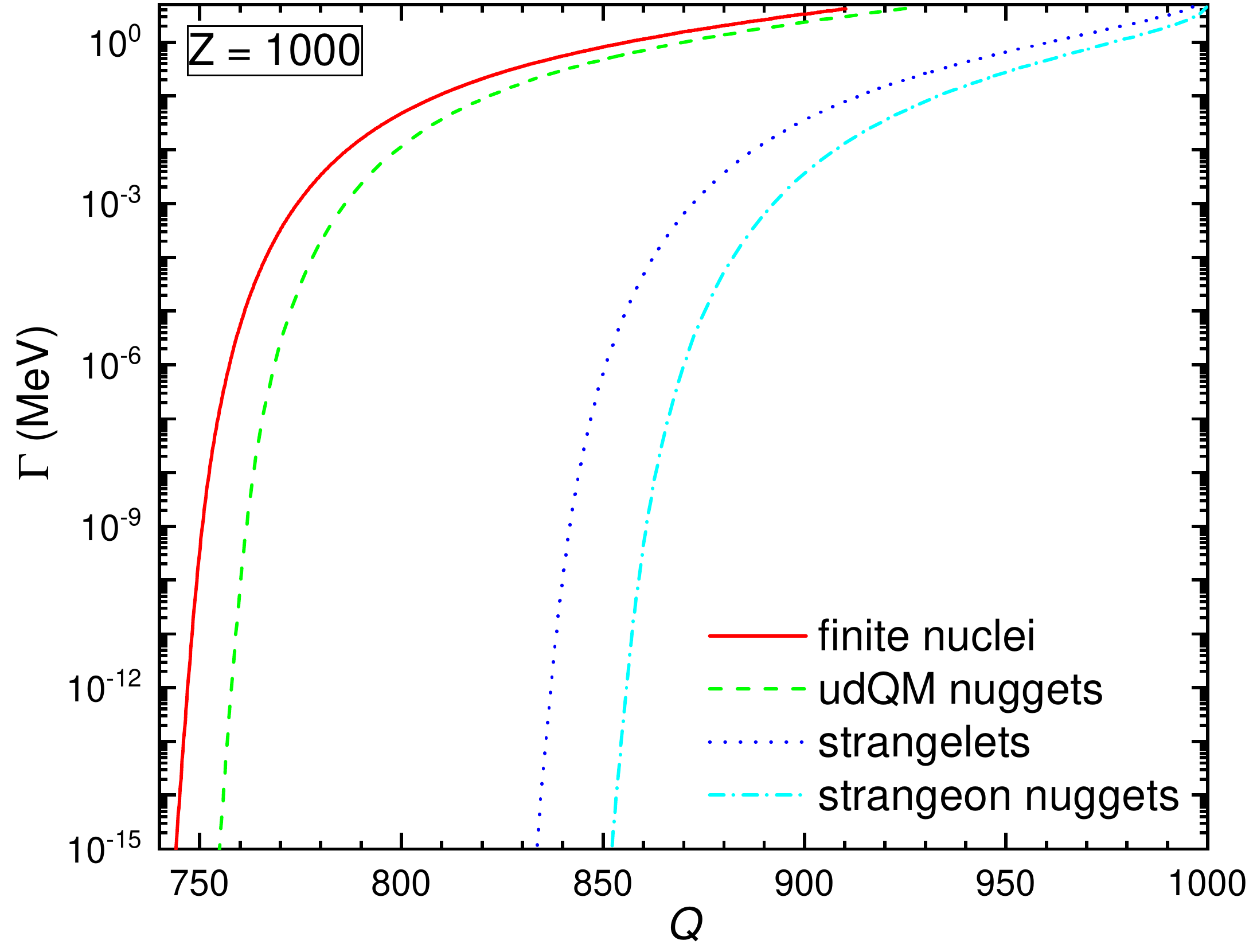}
\end{minipage}
\\
\begin{minipage}[t]{0.49\linewidth}
\centering
\includegraphics[width=\textwidth]{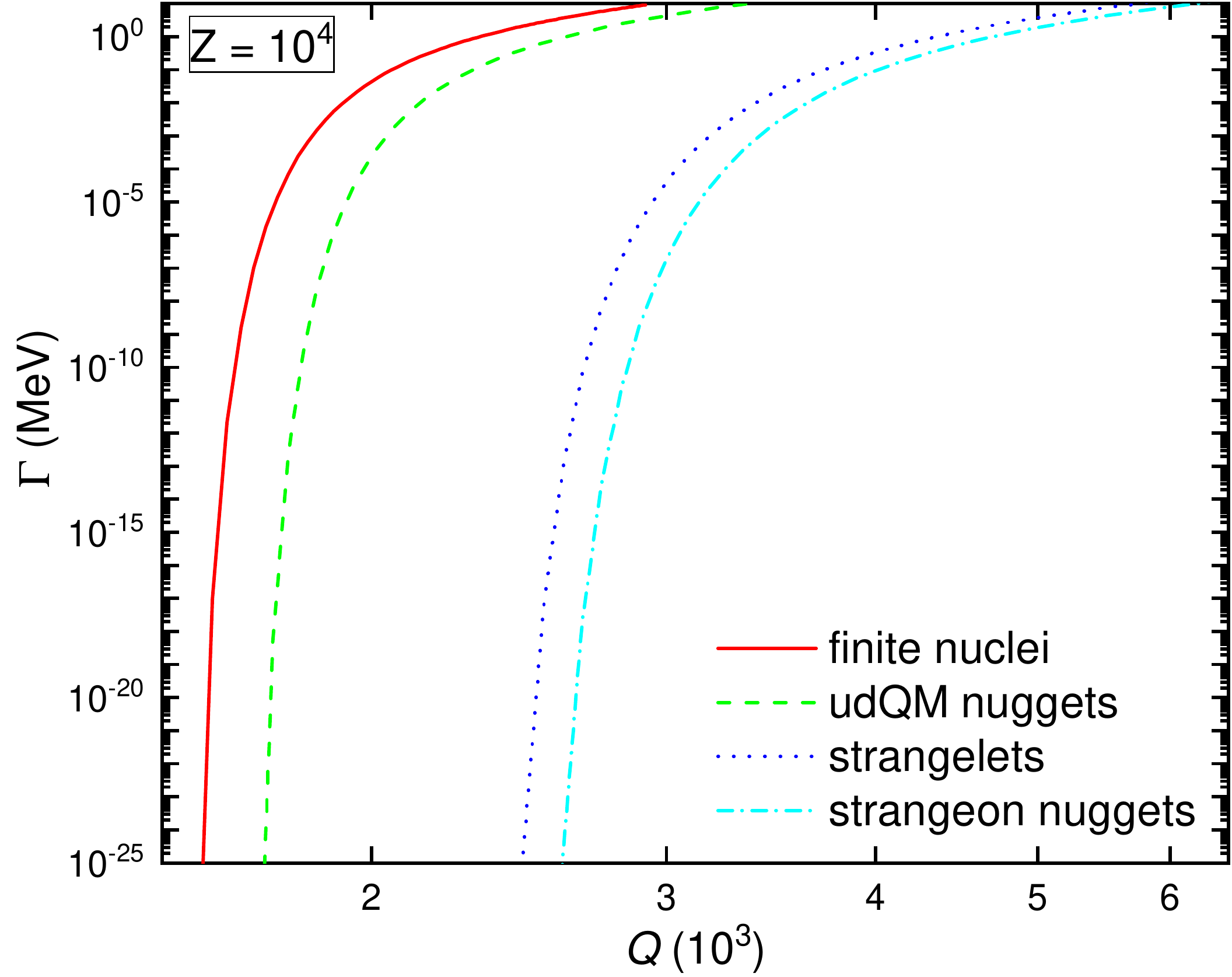}
\end{minipage}%
\hfill
\begin{minipage}[t]{0.505\linewidth}
\centering
\includegraphics[width=1\textwidth]{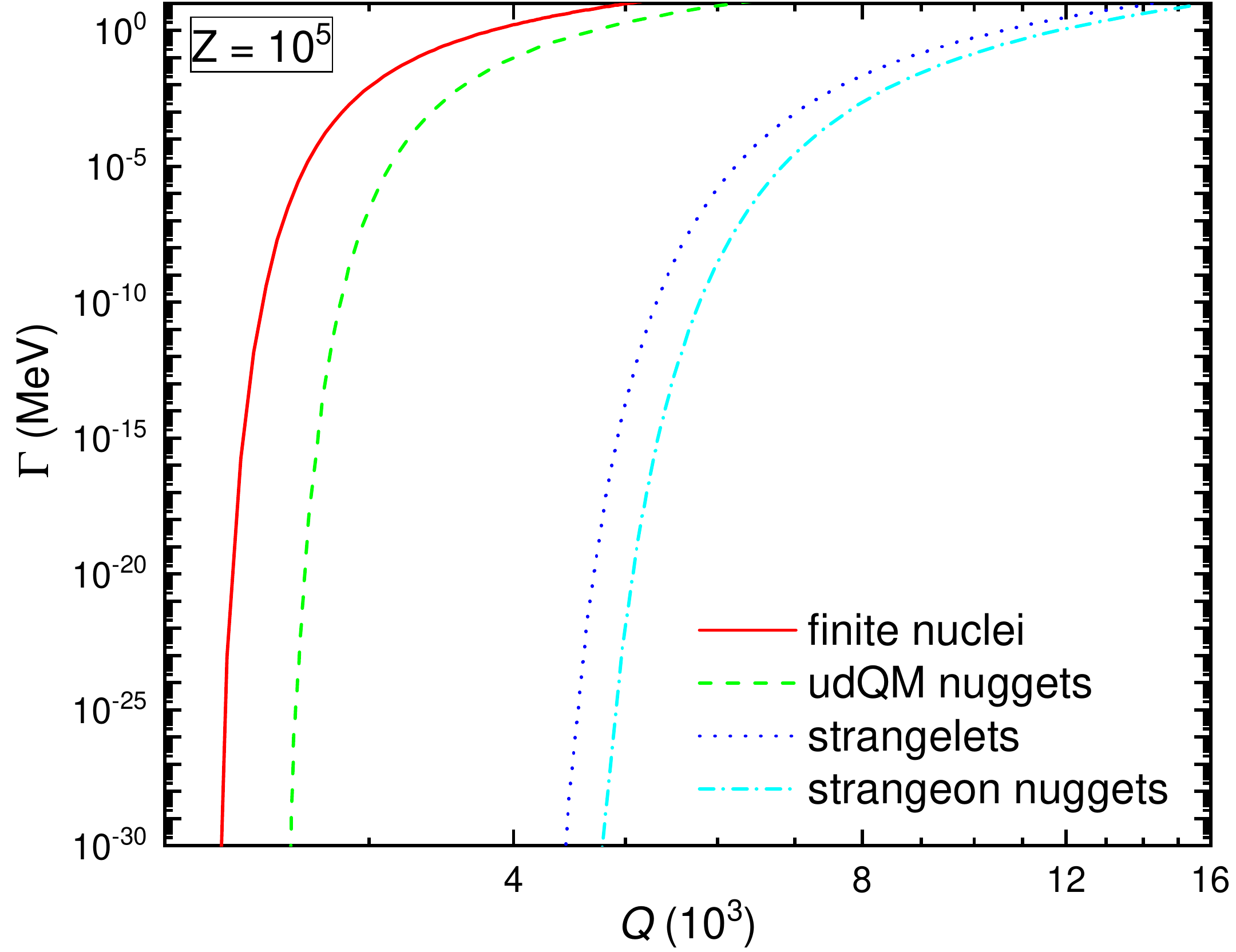}
\end{minipage}
\caption{\label{Fig:NetZ} The decay rates of $e^+e^-$ pair creation for objects with $Z=300$, 1000, $10^4$, and $10^5$, where the total charge $Q$ is fixed at a given $\bar{\mu}_e$.}
\end{figure*}

For a supercritically charged object, the decay rate can then be estimated based on the JWKB approximation~\cite{Kleinert2008_PRD78-025011, Ruffini2010_PR487-1}, i.e.,
\begin{equation}
 \Gamma = \frac{1}{\pi} \int_{V(0)+m_e}^{-m_e} \sum_{l=0}^{l_\mathrm{max}} (2l+1) f(\varepsilon)  P_\mathrm{JWKB}(\varepsilon, l)  \mbox{d}\varepsilon, \label{eq:W_epm}
\end{equation}
with the electron transmission probability at given energy $\varepsilon$ and angular momentum $l$ being
\begin{equation}
P_\mathrm{JWKB} = \exp{\left[2\int_{r_-}^{r_+} \sqrt{\frac{l(l+1)}{r^2} + m_e^2 - \left(\varepsilon+\frac{\alpha Q}{r}\right)^2 } \mbox{d}r \right]}. \label{eq:PWKB}
\end{equation}
Here $f(\varepsilon)$ predicts the empty states of electrons. If the $e^+e^-$ pair creation rate is much smaller than the rate of electron thermalization, we can adopt the Fermi-Dirac distribution of electrons and have
\begin{equation}
f(\varepsilon) =1 - \left[1+\exp{\left(\frac{\varepsilon-\bar{\mu}_e}{T}\right)}\right]^{-1}, \label{eq:Fermi-Dirac}
\end{equation}
where a lower limit $\bar{\mu}_e$ in the integral of Eq.~(\ref{eq:W_epm}) is obtained for zero temperature cases ($T=0$) due to the requirement of Pauli exclusion principle, and the maximum angular momentum is given by $l_\mathrm{max}=\mathrm{Int}\left(\sqrt{\alpha^2 Q^2 + 1/4} - 1/2\right)$. The two real turning points $r_\pm$ are obtained by solving
\begin{equation}
\varepsilon+\frac{\alpha Q}{r_\pm} =  \pm \sqrt{\frac{l(l+1)}{r^2_\pm} + m_e^2},
\end{equation}
which gives
\begin{equation}
r_\pm = -\frac {\alpha Q \varepsilon \pm \sqrt {\alpha^2 Q^2 m_e^2 + l(l+1) \left(\varepsilon^2-m_e^2\right)} }
               {\varepsilon^2-m_e^2}.
\end{equation}
Note that the turning points may become smaller than the electron-vacuum boundary ($r_\pm<R_e$) at $l>0$. The tunneling process for $\varepsilon>\bar{\mu}_e$ is still possible without violating the Pauli exclusion principle. However, the Coulomb potential $V(r)=-{\alpha Q}/{r}$ is not valid at $r<R_e$, since the charge number enclosed within the sphere of radius $r$ becomes larger than $Q$~\cite{Ruffini2011_PLB696-416}. In such cases, $r_+$ may become slightly larger and the transmission probability $P_\mathrm{JWKB}(\varepsilon, l)$ at $l>0$ increases. In this work, for simplicity, we neglect the variation of the Coulomb potential at $r<R_e$. The integral in Eq.~(\ref{eq:PWKB}) can then be obtained with
\begin{equation}
P_\mathrm{JWKB} = \exp{\left[2\pi\sqrt{\alpha^2Q^2 - l(l+1)} +\frac{2\pi\alpha Q\varepsilon}{\sqrt{\varepsilon^2-m_e^2}}  \right]}.
\end{equation}
By taking $l$ as continuum values, the summation in Eq.~(\ref{eq:W_epm}) can be obtained via integration and gives
\begin{eqnarray}
 \Gamma = && \frac{1}{2\pi^3} \left[1+ (2\pi\alpha Q-1) \exp{\left(2\pi\alpha Q\right)}\right] \nonumber\\
          && \times \int_{\bar{\mu}_e}^{-m_e} \exp{\left(\frac{2\pi\alpha Q\varepsilon}{\sqrt{\varepsilon^2-m_e^2}}  \right)}
             \mbox{d}\varepsilon. \label{eq:W_epm1}
\end{eqnarray}

Assuming a constant Coulomb potential inside a core of radius $R$ and net charge number $Z$, the electron distributions at given $\bar{\mu}_e$ can be obtained based on Eqs.~(\ref{eq:pdis}) and (\ref{eq:def0}). Note that for the ultra-relativistic cases with $\varphi(r)\gg m_e$, an analytical solution is obtained for $\varphi(r)$~\cite{Rotondo2008}. The values of $R$ and $Z$ for various types of objects are fixed according to the results indicated in Figs.~\ref{Fig:Charge} and \ref{Fig:Radius}, where $\bar{\mu}_e=-m_e$ was adopted.

The $e^+e^-$ pair production rate is predicted by Eq.~(\ref{eq:W_epm1}), where the total charge number $Q$ is fixed at a given $\bar{\mu}_e$ with $\bar{\mu}_e\leq-m_e$.
In Fig.~\ref{Fig:NetZ} we present our results for supercritically charged nuclei, $ud$QM nuggets, strangelets, and strangeon nuggets with $Z=300$, 1000, $10^4$, and $10^5$. For a supercritically charged object carrying a net charge $Z$, as indicated in Eq.~(\ref{eq:QR_relation}), the total charge number $Q$ decreases from $Z$ as electrons are created and fill in the Fermi sea, while the corresponding positrons leave the system due to Coulomb repulsion. The variation of $Q$ for supercritically charged objects becomes small at $\Gamma\lesssim 10^{-7}$ MeV. This suggests that the $e^+e^-$ pair creation is most effective at $t\lesssim 10^{-15}$ s under the assumption that positions are emitted sequentially and $\Gamma$ does not deviate much from those indicated in Fig.~\ref{Fig:NetZ}. During the merger of binary compact stars, the positron emission due to the release of supercritically charged objects may thus be partially responsible for the short $\gamma$-ray burst~\cite{Goldstein2017_ApJ848-L14, Abbott2017_ApJ848-L13}. For a fixed net charge number $Z$, more $e^+e^-$ pairs are produced by objects with smaller $R$, where $R$ increases in the order of finite nuclei, $ud$QM nuggets, strangelets, and strangeon nuggets. For the superheavy nucleus ${}^{918}$300, to create one $e^+e^-$ pair takes at least a few $10^{-22}$ s with the decay rate on the order of MeV, while longer duration is expected for smaller $Z$~\cite{Simenel2011_EPJWC17-09002}. Note that at small charge numbers such as $Z=300$, the pair creation quickly stops at $\Gamma\lesssim 10^{-7}$ MeV since the Coulomb field is easily screened by electrons with $Q-1<Q_{\bar{\mu}_e=-m_e}$ as indicated in Fig.~\ref{Fig:Charge}. This is not the case for larger objects, where the positron emission tends to last much longer since they possess larger charge numbers.

\begin{figure}
\includegraphics[width=\linewidth]{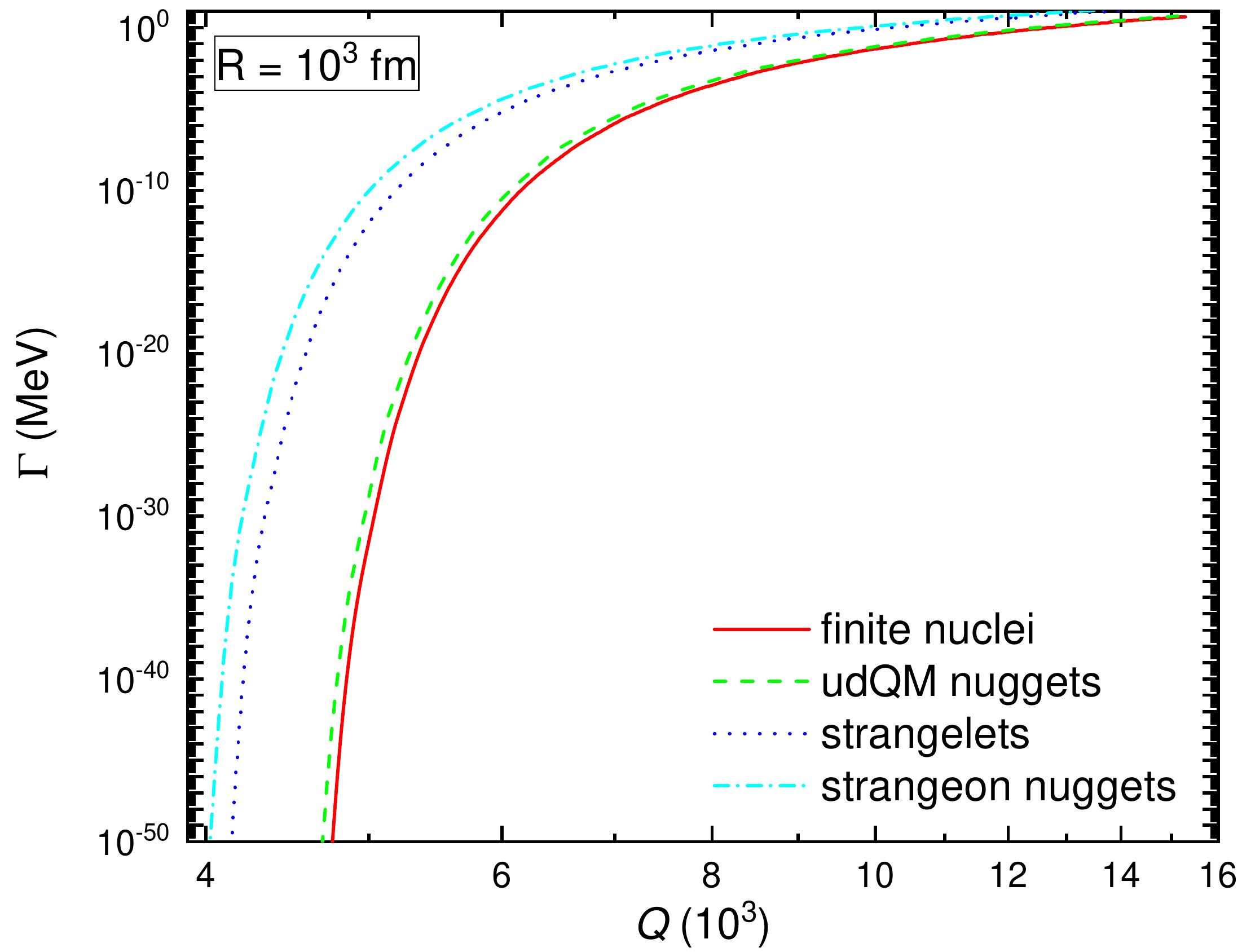}
\caption{\label{Fig:NetR1000} Same as Fig.~\ref{Fig:NetZ} but for supercritically charged objects with $R=1000$ fm.}
\end{figure}

For larger objects, as an example, we consider the cases with $R=1000$ fm, which correspond to the net charge numbers $Z=3.1\times 10^7$, $6.4\times 10^6$, 91698, and 60487 for finite nuclei, $ud$QM nuggets, strangelets, and strangeon nuggets, respectively. The decay rates as functions of the charge number $Q$ are presented in Fig.~\ref{Fig:NetR1000}, which are increasing with $Q$. At $Q\approx 14000$, the obtained decay rates lie in the range of 2--15 MeV, suggesting a fast reduction of $Q$. As $Q$ decreases, the decay rates for $e^+e^-$ pair creation becomes much smaller, i.e., a continued source of positron emission. Comparing with the charge numbers $Q$ ($=Z-N_e$) indicated in Fig.~\ref{Fig:NetZ}, the values obtained here for objects with same radii are close to each other and possess similar decay widths, which is what we have observed in Fig.~\ref{Fig:Charge} for objects with $A\gtrsim 10^8$ or $R\gtrsim 1000$ fm. Meanwhile, similar to static cases, the charge number $Q$ increases with $Z$. As was discussed in Fig.~\ref{Fig:Radius}, a universal relation $Q/R = \left(m_e - \bar{\mu}_e\right)/\alpha$ can be obtained based on Eq.~(\ref{eq:QR_relation}) for very large objects with $R\gtrsim 10^5$ fm or $A\gtrsim 10^{15}$. By substituting this relation into Eq.~(\ref{eq:W_epm1}), the decay rate for objects with $R\gtrsim 10^5$ fm can be determined.

\begin{figure}
\includegraphics[width=\linewidth]{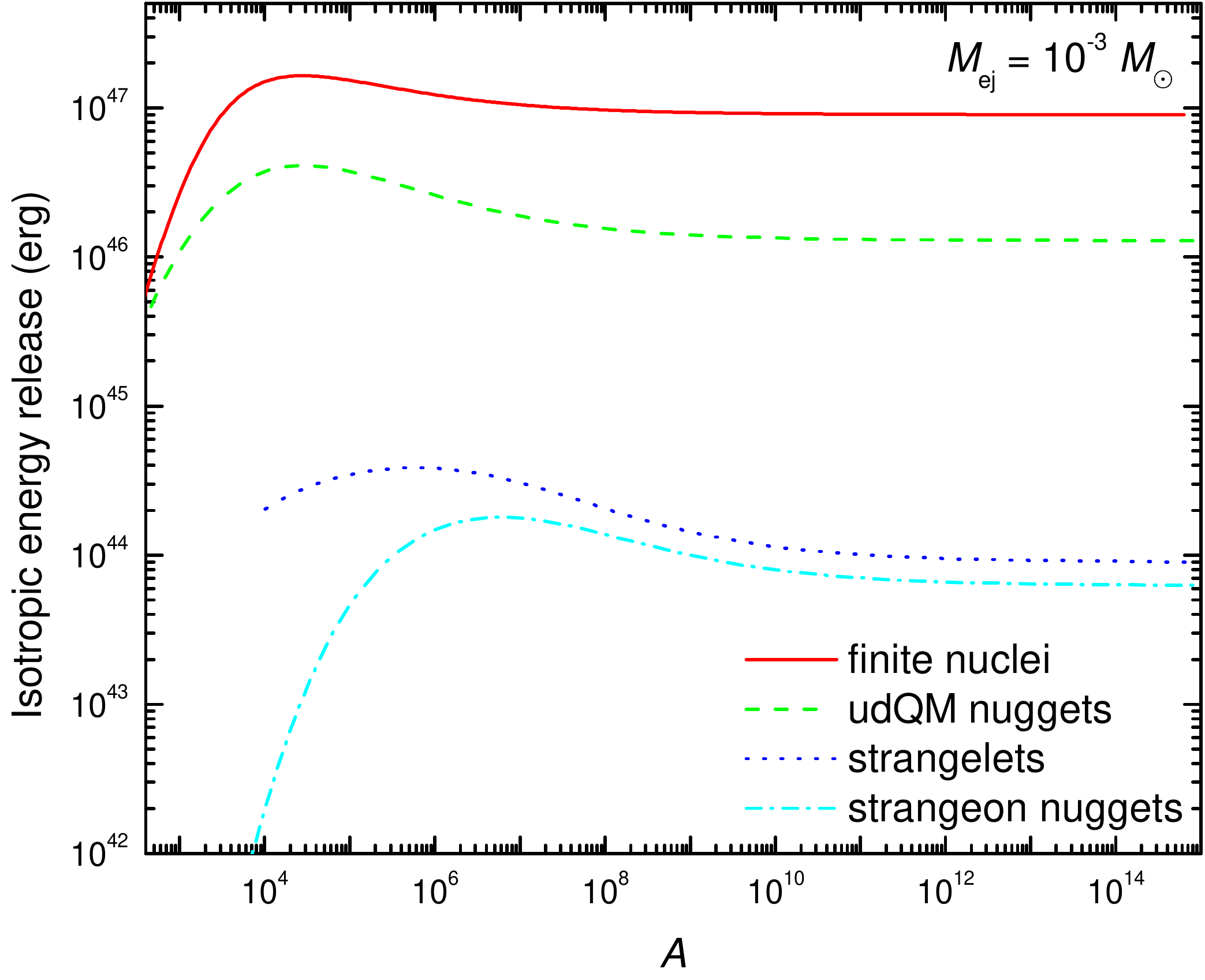}
\caption{\label{Fig:Eiso} Isotropic energy release in $\gamma$-rays via the process of positrons annihilating with electrons.}
\end{figure}

With most $e^+e^-$ pairs created at $t\lesssim 10^{-15}$ s, the maximum number of positrons emitted by supercritically charged objects at $T=0$ can be obtained with $N_{e^+}\approx Z-Q_{\bar{\mu}_e=-m_e}$ based on the charge numbers indicated in Fig.~\ref{Fig:Charge}. In Fig.~\ref{Fig:Eiso} a rough estimation on the energy release (${\cal E} \approx 2 m_e N_{e^+} M_\mathrm{ej}/M_A$) of positron annihilation during the merger of binary compact stars is presented, where we have assumed $M_\mathrm{ej}=0.001\ M_\odot$ for the total mass of ejected objects with baryon number $A$ and mass $M_A$ as determined by Eq.~(\ref{eq:mass}). It should be mentioned that the ejected mass ($\sim10^{-5}$-$10^{-2}\ M_\odot$) and its composition depend on binary parameters and the equation of state of dense stellar matter~\cite{Perez-Garcia2013_ApJ768-145, Radice2018_ApJ869-130}, which are likely to deviate from our current assumption. The obtained isotropic energy release for the ejected superheavy nuclei and $ud$QM nuggets are comparable with the estimated value $(3.1 \pm 0.7)\times 10^{46}$ erg of GRB 170817A~\cite{Goldstein2017_ApJ848-L14, Abbott2017_ApJ848-L13}, while smaller values for strangelets and strangeon nuggets are obtained. With such a substantial amount of $e^+e^-$ pairs produced within a compact region and a short period of time, drastic collisions among electrons, positrons, photons, various types of particles, and supercritically charged objects take place, which become optically thick to $\gamma$-rays and would reach thermal equilibrium if the thermalization time is shorter than the escape time. A reduction of photon peak energy from positron annihilation (511 keV) is thus expected, e.g., a blackbody spectrum with a high-energy tail~\cite{Aksenov2004_ApJ609-363}. Note that there is a 1.7 second delay between the trigger times of the gravitational-wave signal GW170817 and the $\gamma$-ray burst GRB 170817A~\cite{Goldstein2017_ApJ848-L14, Abbott2017_ApJ848-L13}, which may be attributed to two main reasons~\cite{Abbott2017_ApJ848-L13}: 1. the intrinsic delay between the moment of binary coalescence and the production of an emitting region, e.g., the time it takes for the ionization process to take effects and/or the launching of a relativistic jet;  2. the time elapsed for the emitting region to become transparent to $\gamma$-rays, e.g., the required time for the fireball to expand and become optically thin to $\gamma$-rays and/or the propagation of the jet to break out of the dense gaseous environment.

It is worth mentioning that the temperature can be as high as $T\approx 50$ MeV during the merger of a binary system, e.g., those indicated in Ref.~\cite{Most2019_PRL122-061101}. The thermal electron-positron pairs will thus be produced abundantly, where the number density of positrons can be fixed by $n_{e^+} \approx 2.378\times 10^{-8}T^3$ with $n_{e^+}$ in $\mathrm{fm}^{-3}$ and $T\ (\gtrsim m_e)$ in MeV. If we suppose there is a heated spherical region ($T = 50$ MeV) with a radius $\sim$2 km in the ejecta~\cite{Most2019_PRL122-061101}, the corresponding energy stored within the rest mass of $e^+e^-$ pairs is ${\cal E} \approx 1.6\times10^{47}$ erg. This value may become larger if we consider the other regions of ejecta, though most of the energy may be converted into kinetic energy as ejecta expands~\cite{Abbott2017_ApJ848-L13}. Meanwhile, we should mention there may be other important energy sources, e.g., the thermonuclear reactions, the thermal radiation such as the outflowing $\nu\bar{\nu}$~\cite{Shibata2006_PRL96-031102} and/or $e^+e^-$~\cite{Usov1998_PRL80-230} fluxes, the decay of strangelets~\cite{Bucciantini2019} and strangeon nuggets~\cite{Lai2018_RAA18-024}, etc. In such cases, the energy release in $\gamma$-rays during the merger of binary strange stars or strangeon stars can be attributed to those processes instead of positron emissions from strangelets or strangeon nuggets.

A substantial amount of positrons and supercritically charged objects may finally escape the binary system, which later create the 511 keV continuum emission observed in the Galaxy via positronium decay~\cite{Weidenspointner2006_AA450-1013, Prantzos2011_RMP83-1001}. In fact, it was shown that the observed positron annihilation mainly comes from the bulge with a large bulge-to-disk ratio around 1.4~\cite{Prantzos2011_RMP83-1001}, which seems to correlate with the distribution of binary systems in the Milky Way. Such kinds of correlations have recently been adopted as tracers of binary neutron star mergers~\cite{Fuller2019_PRL122-121101}. Meanwhile, before the emission of positrons, the $e^+e^-$ pairs produced around the surfaces of supercritically charged objects would oscillate with alternating electric field for a short time, and emit electromagnetic radiations with a characteristic frequency around 4 keV~\cite{Han2010_PLB691-99}. We suspect these radiations are actually responsible for the narrow faint emission lines around 3.5, 8.7, 9.4 and 10.1 keV observed in the Milky Way center, nearby galaxies and galaxy clusters~\cite{Iakubovskyi2016_AASP6-3, Conlon2017_PRD96-123009}.

\section{\label{sec:con}Conclusion}
We study the properties of finite-sized objects that are heavier than the currently known nuclei, i.e., superheavy nuclei, $ud$QM nuggets, strangelets, and strangeon nuggets. The structures of those objects are obtained based on the UDS model~\cite{Xia2016_SciBull61-172, Xia2016_SciSinPMA46-012021_E, Xia2016_PRD93-085025, Xia2017_NPB916-669}, where the Thomas-Fermi approximation is adopted. The local properties of nuclear matter, $ud$ quark matter, strange quark matter, and strangeon matter are determined by expanding the energy per baryon to the second order, while a surface tension is introduced for the hadron/quark-vacuum interface. The parameters are fixed by reproducing the masses and charge properties of $\beta$-stable nuclei~\cite{Audi2017_CPC41-030001, Huang2017_CPC41-030002, Wang2017_CPC41-030003}, $ud$QM nuggets~\cite{Holdom2018_PRL120-222001}, large strangelets~\cite{Xia2017_NPB916-669}, and strangeon matter~\cite{Lai2009_MMRAS398-L31}.

Comparing with the most stable nucleus ${}^{56}$Fe, $ud$QM nuggets, strangelets, and strangeon nuggets are more stable at $A>A_\mathrm{crit}$ with $A_\mathrm{crit} \approx 315$, $5\times10^4$, and $1.2\times10^8$, respectively. The masses of finite nuclei and $ud$QM nuggets become similar at $A\approx 266$, which increases the possibility in synthesizing $ud$QM nuggets via heavy ion collisions. The stability of those objects is investigated by examining their chemical potentials, where we have obtained a maximum baryon number for superheavy elements with $A_\mathrm{max}\approx 965$, and minimum baryon numbers $A_\mathrm{min}\approx 39$, 433, and $2.7\times10^5$ for $ud$QM nuggets, strangelets, and strangeon nuggets that are stable against neutron emission. The charge properties of those objects are obtained, where the net charge fraction ($Z/A$) vary smoothly from 0.5, 0.5, 0.1, and 0.0063 ($A\lesssim 100$) to 0.047, 0.0064, $4.6\times 10^{-5}$, and $3.2\times 10^{-5}$ ($A\gtrsim 10^9$) for finite nuclei, $ud$QM nuggets, strangelets, and strangeon nuggets, respectively. For objects with large enough net charge numbers $Z\geq Z_\mathrm{crit}$, $e^+e^-$ pair creation inevitably starts, where $Z_\mathrm{crit} = 163$, 177, 192, and 212 for $ud$QM nuggets (${}^{609}163$), finite nuclei (${}^{480}$177), strangelets (${}^{16285}$192), and strangeon nuggets (${}^{90796}$212), respectively. The maximum charge numbers that are stable against $e^+e^-$ pair creation are investigated, which increase with $Z$ and are converging at $R\gtrsim 1000$ fm or $A\gtrsim 10^8$ for different types of objects. A universal relation $Q/R_e = \left(m_e - \bar{\mu}_e\right)/\alpha$ is obtained at given $\bar{\mu}_e$, where $Q$ the charge and $R_e$ the radius of electron cloud. The maximum charge can be obtained by taking $\bar{\mu}_e=-m_e$. At $R\gtrsim 10^5$ fm or $A\gtrsim 10^{15}$, $R \approx R_e$ and the universal charge radius relation is obtained with $Q = 0.71 R$, which is consistent with those predicted in Ref.~\cite{Madsen2008_PRL100-151102}.

For supercritically charged objects, the decay rate for $e^+e^-$ pair production is estimated based on the JWKB approximation~\cite{Kleinert2008_PRD78-025011, Ruffini2010_PR487-1}. It is found that most positrons are emitted at $t\lesssim 10^{-15}$ s, which should be partially responsible for the short $\gamma$-ray burst due to the release of supercritically charged objects during the merger of binary compact stars~\cite{Goldstein2017_ApJ848-L14, Abbott2017_ApJ848-L13}. For the superheavy nucleus ${}^{918}$300, to create one $e^+e^-$ pair requires at least few $10^{-22}$ s, while longer duration is expected for smaller $Z$. The $e^+e^-$ pair creation for small objects ($Z=300$) quickly stops due to the screening effects of electrons. For larger objects, positron emission last much longer, which may be responsible for the 511 keV emission from positron annihilation in the Galaxy~\cite{Weidenspointner2006_AA450-1013, Prantzos2011_RMP83-1001} as well as the narrow faint emission lines in X-ray spectra observed in the Milky Way center, nearby galaxies and galaxy clusters~\cite{Iakubovskyi2016_AASP6-3, Conlon2017_PRD96-123009}.

Finally, it is worth mentioning that the temperature of newly created supercritically charged objects may reach up to $\sim$50 MeV during the merger of a binary system~\cite{Baiotti2019_PPNP109-103714}. In such cases, the rate of $e^+e^-$ pair creation becomes much larger since the electronic states with $\varepsilon<\bar{\mu}_e$ may not be completely occupied as predicted in Eq.~(\ref{eq:Fermi-Dirac}). The thermal ionization should also be considered, where bound electrons are excited to the continuum of free electron states so that the charge $Q$ of those objects is increased. In fact, the emission of positrons due to $e^+e^-$ pair creation combined with the evaporation of thermalized electrons was shown to create an outflowing plasma of $\sim$$10^{51}$ ergs/s on strange stars' surfaces with $T\approx10^{11}$ K~\cite{Usov1998_PRL80-230}. Meanwhile, the environment of these objects created during the merger of a binary system may be filled with $e^+e^-$ plasma, which could reduce $Q$ by capturing the surrounding electrons. In such cases, to determine the final state of those charged objects, more detailed studies on the evolution of $Q$ with $e^+e^-$ pair creation, thermal ionization, and electron capturing combined with the time evolution of their surrounding environment are necessary, which is intended in our future works. Due to the requirement of charge conservation, same amount of electrons $N_{e}=Q$ are ejected from the charged object. Some of the electrons will recombine with the positively charged objects, or experience a positronium decay with the positrons emitted by supercritically charged objects, while the rest of them forms a $e^+e^-$ plasma or trapped along magnetic field lines and emit synchrotron radiation. All of which are expected to contribute to the electromagnetic signal of the short $\gamma$-ray bursts. Nevertheless, we do not know for sure how many of those supercritically charged objects are created or the exact charge number $Q$ they carry, in which case a detailed dynamical simulation on those processes needs to be carried out.

\section*{ACKNOWLEDGMENTS}
C.J.X. would like to thank Prof. Bao-An Li for fruitful discussions. This work was supported by National Natural Science Foundation of China (Grants No.~11705163, No.~11875052, No.~11673002,  No.~11525524, No.~11621131001, No.~11947302, and No.~11961141004), Ningbo Natural Science Foundation (Grant No.~2019A610066), the National Key R\&D Program of China (Grant No.~2018YFA0404402), the Key Research Program of Frontier Sciences of Chinese Academy of Sciences (No.~QYZDB-SSWSYS013), and the Strategic Priority Research Program of Chinese Academy of Sciences (Grant No.~XDB34010000). The computation for this work was supported by the HPC Cluster of ITP-CAS and the Supercomputing Center, Computer Network Information Center of Chinese Academy of Sciences.


\newpage

%

\end{document}